%Paper: astro-ph/9411059
%From: yi@cfata3.harvard.edu (insu yi)
%Date: Tue, 15 Nov 94 18:11:53 -0500
%Date (revised): Tue, 15 Nov 94 18:54:11 -0500
%Date (revised): Sat, 29 Jul 95 17:47:14 -0400

%%%%%%%%%%%%%%%%%%%%%%%%%%%%%%%%%%%%%%%%%%%%%%%%%%%%%%%%%%%%%%%%%%%%%%%
%Advection-Dominated Accretion: Underfed Black Holes and Neutron Stars
%Ramesh Narayan and Insu Yi
%To Appear in Astrophysical Journal Oct. 20, 1995
%Revised, Expanded, Final Version
%%%%%%%%%%%%%%%%%%%%%%%%%%%%%%%%%%%%%%%%%%%%%%%%%%%%%%%%%%%%%%%%%%%%%%%
\magnification=\magstep1
\def\ep{\epsilon}
\def\epp{\epsilon^{\prime}}
\def\o{\over}
\def\md{\dot m}
\def\mda{\left({\dot m\over\alpha^2}\right)}
\def\rs{r}
\def\sun{_{\odot}}
\def\ssp{\baselineskip=11pt plus 1pt minus 1pt}

\spaceskip=0.4em plus 0.15em minus 0.15em
\xspaceskip=0.5em
\hsize=17 true cm
\hoffset=0 true cm
\vsize=22 true cm

\def\ref{\par\noindent\hangindent 20pt}
\def\sles{\lower2pt\hbox{$\buildrel {\scriptstyle <}
   \over {\scriptstyle\sim}$}}
%insert forced spacing before and after \sles, \sgreat within $--$

\def\sgreat{\lower2pt\hbox{$\buildrel {\scriptstyle >}
   \over {\scriptstyle\sim}$}}
%needs forced spacing, as above

\def\lapprox{\lower2pt\hbox{$\buildrel \lower2pt\hbox{${\scriptstyle<}$}
   \over {\scriptstyle\approx}$}}
%needs forced spacing, as above

\def\gapprox{\lower2pt\hbox{$\buildrel \lower2pt\hbox{${\scriptstyle>}$}
   \over {\scriptstyle\approx}$}}
%ditto

\def\both{\lower2pt \hbox{$\buildrel {\leftarrow} \over {\rightarrow}$}}
%ditto

 %requires $...$ in text

%\dsp
\ssp
\centerline{\bf Advection-Dominated Accretion: Underfed Black Holes and
Neutron Stars}
\bigskip
\centerline{Ramesh Narayan and Insu Yi}
\bigskip
\centerline{Harvard-Smithsonian Center for Astrophysics}
\centerline{60 Garden Street, Cambridge, MA 02138}
\vskip .4in

\noindent{\bf Abstract}
\bigskip

We describe new optically thin solutions for rotating accretion flows
around black holes and neutron stars.  These solutions are
advection-dominated, so that most of the viscously dissipated energy
is advected radially with the flow.  We model the accreting gas as a
two-temperature plasma and include cooling by bremsstrahlung,
synchrotron, and Comptonization.  We obtain electron temperatures
$T_e\sim 10^{8.5}-10^{10}$K.

The new solutions are present only for mass accretion rates $\dot M$
less than a critical rate $\dot M_{crit}$ which we calculate as a function
of radius $R$ and viscosity parameter $\alpha$.  For $\dot M<\dot
M_{crit}$ we show that there are three equilibrium branches of
solutions.  One of the branches corresponds to a cool optically thick
flow which is the well-known thin disk solution of Shakura
\& Sunyaev (1973).  Another branch corresponds to a hot optically thin
flow, discovered originally by Shapiro, Lightman \& Eardley (1976,
SLE).  This solution is thermally unstable.  The third branch
corresponds to our new advection-dominated solution.  This solution is
hotter and more optically thin than the SLE solution, but is viscously
and thermally stable.  It is related to the ion torus model of Rees et
al. (1982) and may potentially explain the hard X-ray and $\gamma$-ray
emission from X-ray binaries and active galactic nuclei.

For $\dot M<\dot M_{crit}$, our work suggests that an accretion flow
can choose between two distinct states, namely the thin disk solution
and the new advection-dominated solution, both of which are apparently
stable.  We argue that, in certain circumstances, it is only the
latter solution that is truly stable, and that a thin disk will
spontaneously evaporate and convert itself into an advection-dominated
flow.  Even for $\dot M>\dot M_{crit}$ we suggest that a thin disk may
evaporate partially so that a fraction of the accretion occurs via an
advection-dominated hot corona.  If these ideas are correct, then
optically thin advection-dominated flows must be very widespread,
possibly the most common form of accretion in black holes accreting at
sub-Eddington $\dot M$.

Our calculations indicate that advection-dominated accretion on black
holes differs considerably from similar flows around neutron stars.
The crucial physical difference, which has been previously mentioned
in the literature, is that in the former the advected energy is lost
into the hole whereas in the latter it is thermalized and reradiated
at the stellar surface, thereby providing soft photons which can
Compton-cool the accreting gas.  We obtain $\dot M_{crit}
\sim\alpha^2\dot M_{Edd}$ for accreting black holes, independent
of the black hole mass, whereas it is $\sim0.1\alpha^2\dot M_{Edd}$
for neutron stars.  Advection-dominated accretion is therefore more
likely to occur in accreting black holes, and these systems will be
underluminous for their $\dot M$ because the bulk of the energy is
advected into the hole rather than being radiated.  We find that $T_e$
in accreting black hole flows rises up to $\sim 10^9-10^{10}$K,
compared to $T_e\sim 10^{8.5}-10^{9}$K in neutron star systems.
Spectra of accreting black holes are therefore expected to be harder
than those of accreting neutron stars.  Pair effects are also more
likely in black hole systems, though only at higher $\dot M$ than
those we consider.

\bigskip\bigskip\noindent
Subject headings: accretion, accretion disks, black hole physics,
galaxies: nuclei, quasars, hydrodynamics, stars: neutron, X-rays:
stars

\vfill\eject
\noindent{\bf 1. Introduction}
\bigskip

One of the cornerstones of the theory of accretion disks is the model
of thin disks developed by Shakura \& Sunyaev (1973), Novikov \&
Thorne (1973), and Lynden-Bell \& Pringle (1974) (see Frank, King \&
Raine 1992 for a review).  This model, which has been widely applied
to many accreting systems, is based on the assumption that the
accreting material cools efficiently so that all the energy which is
released through viscosity is radiated locally.  As a consequence,
the accreting gas is much cooler than the local virial temperature,
and therefore the orbiting material has a vertical thickness which is
much smaller than the radius.

What happens if the cooling in an accretion flow is not efficient?
Clearly, if the cooling is unable to keep up with the heating, then a
part of the viscously released energy will have to be advected with
the accreting gas as stored entropy.  We may then expect the gas to be
driven to a higher temperature, which will lead to a vertical
thickening of the disk.  Pressure forces also will become important
and will modify the dynamics; for instance, the rotation is likely to
be sub-Keplerian.  Accretion flows with some of these properties have
been considered from time to time in various contexts (e.g. Begelman
1978, Liang \& Thompson 1980, Paczy\'nski \& Wiita 1980, Phinney 1981,
Rees et al. 1982, Begelman \& Meier 1982, Liang 1988, Eggum, Coroniti
\& Katz 1988, Abramowicz et al. 1988, Narayan \& Yi 1994, 1995,
hereafter Papers 1, 2).

The basic dynamical equations of accretion with inclusion of entropy
advection have been written by various workers, notably Paczy\'nski
\& Bisnovatyi-Kogan (1981) and Muchotrzeb \& Paczy\'nski (1982) who
derived a particular set of equations using a simplifying
height-integration approximation.  The formalism was developed further
by Abramowicz et al. (1988), who named the approach the ``slim disk''
model, and carried out a comprehensive analysis of accretion flows
around black holes with mass accretion rates $\dot M$ comparable to or
greater than the Eddington rate. These authors discovered a new
solution branch at super-Eddington $\dot M$ where the optical depth of
the accreting gas is so high that the diffusion time becomes longer
than the viscous time (see Begelman 1978 for an early discussion of
this kind of radiation trapping).  Consequently, the gas is unable to
cool and instead advects the dissipated energy.  The new
advection-dominated solution of Abramowicz et al. (1988) is both
thermally and viscously stable.

Following on the work of Abramowicz et al. (1988), the slim disk model
has been applied to the study of thermal and viscous instabilities and
limit cycles in optically thick accretion disks (e.g.  Honma,
Matsumoto \& Kato 1991, Wallinder 1991, Chen \& Taam 1993).  It has
also been used successfully in the modeling of boundary layers
(Narayan \& Popham 1993, Popham et al. 1993, Popham \& Narayan 1995).

Beginning with the pioneering work of Shapiro, Lightman \& Eardley
(1976, hereafter SLE), investigations of black hole accretion disks at
lower $\dot M$ have focused on a class of optically thin solutions
where the gas is significantly hotter than the local Shakura-Sunyaev
solution.  The accreting plasma in these solutions is two-temperature,
with the ions being significantly hotter than the electrons.  Since
the gas is optically thin, the cooling occurs through bremsstrahlung
or via Comptonization of the bremsstrahlung photons and other soft
photons that may be produced locally.  These hot disk models have been
applied to the various states of Cyg X-1 (SLE 1976, Melia \& Misra
1994), while a related model, the ``ion torus'' (Rees et al. 1982),
has been applied to active galactic nuclei (AGN) at low $\dot M$.
Recently, Wandel \& Liang (1991) and Luo \& Liang (1994) introduced a
phenomenological bridging formula in the radiative transfer equations
to display the topological relationship between the SLE hot solution
and the Shakura-Sunyaev thin disk.  Kusunose \& Takahara (1989)
considered the effects of electron-positron pairs, magnetic fields and
cyclotron emission.

A major attraction of the SLE branch of solutions is that the plasma
attains electron temperatures $T_e\sim10^9$ K, so that these solutions
have the promise of being able to explain the hard X-ray and
$\gamma$-ray emission seen in X-ray binaries and AGN.  Unfortunately,
the solutions are thermally unstable (though they are viscously
stable).  The thermal instability arises because the accreting gas is
optically thin so that the cooling efficiency via bremsstrahlung
decreases with decreasing density (Pringle, Rees \& Pacholczyk 1973,
Piran 1978).  Because of this feature, if an equilibrium SLE flow is
perturbed to a slightly higher temperature, its density goes down and
the rate of cooling decreases.  This causes the gas to heat up
further, leading to a runaway thermal instability.  Kusunose \&
Takahara (1989) showed that the inclusion of pairs in the physics of
the hot plasma does not remove the instability.

What is the fate of an optically thin flow which encounters the
thermal instability?  In analogy with the Abramowicz et al. (1988)
work on optically thick flows described above, one might suspect that
optically thin flows may again have a hitherto undiscovered solution
in which entropy advection dominates.  If such an advection-dominated
solution were present, it would be even hotter and more optically thin
than the SLE solution, and it is likely to be stable.  Surprisingly,
this point does not seem to have been fully appreciated until very
recently.  There is a suggestion of these ideas in the ion torus paper
of Rees et al. (1982).  However, the first real hint that advection
can introduce stability in optically thin flows came with the work of
Narayan \& Popham (1993), who studied boundary layers in cataclysmic
variables (CVs) and found that at low $\dot M$ the boundary layer
becomes optically thin and thermally unstable.  However, the
instability does not affect the overall flow significantly, since the
accreting material just switches to a local advection-dominated state
in the optically thin zone, a point emphasized more recently by
Abramowicz et al. (1995).

Following up on the Narayan \& Popham (1993) work, the present authors
investigated in Papers 1 and 2 the general properties of
advection-dominated flows.  In these papers we derived self-similar
solutions, both of the slim disk equations (Paper 1) and of a more
general non-height-integrated set of equations (Paper 2).  In earlier
work, Spruit et al. (1987) considered self-similar height-integrated
solutions in the context of the spiral-shock wave driven instability
and found that for low values of the ratio of specific heats accretion
without radiative losses is possible.  The solutions in Papers 1 and 2
give us quantitative estimates of the angular velocity, radial
velocity, pressure, temperature, etc.  of an advection-dominated
accretion flow as functions of a few basic parameters, such as the
ratio of specific heats $\gamma$ of the gas and the viscosity
parameter $\alpha$.  More importantly, the solutions reveal that
advection-dominated flows have qualitatively very different
properties than standard thin accretion disks: (1) advection-dominated
flows are almost spherical in their morphology (not at all disk-like);
(2) they do not have empty funnels (as some workers have suggested
they might); (3) they rotate quite slowly compared to Keplerian
(tending to zero rotation as $\gamma\rightarrow5/3$); (4) the radial
flow velocities are comparable to the free fall velocity (therefore
accretion occurs much more rapidly than in a thin disk); (5) the flows
are convectively unstable (which would contribute at least partially
to the viscosity); (6) they often have a positive Bernoulli constant
(which may lead to the formation of jets and outflows); and (7) by
definition, advection-dominated flows radiate much less efficiently
than thin disks, and therefore are ultra-dim for their accretion rate
(Phinney 1981, Rees et al. 1982).

In Papers 1 and 2 we discussed the two relevant limits under which an
accretion flow becomes advection-dominated, namely at very high
accretion rates, where the optical depth of the gas is very large
(Begelman 1978, Abramowicz et al. 1988), and at low accretion rates,
where the optical depth of the gas becomes significantly less than
unity (Rees et al. 1982, Narayan \& Popham 1993, Abramowicz et
al. 1995).  However, lacking a detailed model, we could not determine
whether or not advection-dominated accretion is truly relevant to real
systems in nature.  We investigate this question in this paper for the
low $\dot M$ optically thin regime mentioned above. In \S2 we
summarize the basic results of Papers 1 and 2 and write down the
self-similar solutions derived in these papers.  The calculations
presented in later sections are based on these solutions.  Then, in
\S3, we set up a fairly detailed set of equations to describe the
heating and cooling of a two-temperature plasma in an accretion flow.
We include a number of cooling processes, such as bremsstrahlung,
synchrotron, and Compton cooling, and model energy transfer from the
ions to the electrons through Coulomb scattering.  To model these
processes we make use of standard formulae taken from the literature.

We present the main results of the paper in \S4.  We show that the
equations we develop in \S\S2,3 permit three branches of solutions:
the first branch is the standard Shakura-Sunyaev thin accretion disk
solution, the second branch is a hot thermally unstable optically thin
solution which is none other than the hot SLE solution, while the
third branch is a new even hotter and more optically thin solution.
The third solution is advection-dominated, as we might have suspected,
and is thermally stable, again as expected.  We describe the
properties of the new advection-dominated solutions and determine the
regions of the $\dot MR$ plane over which this branch of solutions is
present.  Some of these results were discussed by Rees et al. (1982)
using a more qualitative analysis.  In the calculations presented
here, we model various kinds of central stars, focusing in particular
on three cases, viz.  a $10M_\odot$ black hole, a $10^8M_\odot$ black
hole, and a $1.4M_\odot$ neutron star.  The first and third cases are
relevant to galactic X-ray binaries at low accretion rates and the
second corresponds to underfed active galactic nuclei.  In two recent
papers, Abramowicz et al. (1995) and Chen (1995) have discussed
several of the issues discussed in \S4, but using a simplified
single-temperature model and including only free-free cooling (see
\S4.1).

One of the results that comes out of our calculations is that for a
wide range of conditions both the Shakura-Sunyaev thin disk solution
and the new advection-dominated solution are present, both solutions
being stable within the slim disk approximation.  Which of the two
solution branches will an actual accretion flow choose?  We discuss
this issue in \S5 and make heuristic arguments to suggest that perhaps
it is the advection-dominated branch which is often preferred.  We
conclude in \S6 with a summary of the results and discuss some of the
implications.

The Appendices discuss two technical points.  In Appendix A, we
investigate whether the non-thermal coupling of ions and electrons
described by Begelman \& Chiueh (1988) might invalidate our assumption
of a purely Coulomb-coupled two-temperature plasma.  In Appendix B, we
discuss the effect of radiative viscosity.

Readers who wish to see the main results and are not interested
in the technical details of the model may wish to proceed directly to
\S4 at this point, or perhaps even to \S6 which gives a comprehensive
summary.

Before diving into the technical details in the next two sections, it
is useful to mention the differences between this study and some
previous works.  The major feature of this work (and also the paper by
Abramowicz et al. 1995) is that we systematically include entropy
advection in the dynamics of the flow.  Thereby, we are able to study
optically-thin, two-temperature, advection-dominated flows, which up
until now have been hardly considered at all, and if at all only in a
qualitative manner.  Our use of self-similar solutions (Papers 1, 2,
Begelman \& Meier 1982, Spruit et al. 1987) make the dynamical
structure of our accretion flows unambiguous, whereas previous studies
of thick disks (e.g. Paczy\'nski \& Wiita 1980) and ion tori
(e.g. Rees et al. 1982) left some dynamical quantities unspecified.
As a result, our model allows us to determine self-consistently the
electron densities and temperatures, the magnetic field densities, and
the resulting radiative processes.  There are two technical
differences between our calculations and those of SLE (see also
Eardley et al. 1978).  One is that by including advection we find two
different hot solutions, one of which is stable, whereas without
advection SLE (and later workers) could obtain only one solution, an
unstable one.  Secondly, most of SLE's calculations involved
unsaturated Comptonization of externally supplied soft photons whereas
we have an internal supply of soft photons via synchro-cyclotron
emission by the thermal electrons (e.g. Rees et al. 1982).  Also, we
use a Comptonization formula which automatically works both in the
unsaturated and saturated limits.

\vfill\eject
\noindent
{\bf 2. Advection-Dominated Accretion Flows}
\bigskip\bigskip
\noindent{\bf 2.1. Self-Similar Advection-Dominated Flows}
\bigskip

In Paper 1, we derived the following self-similar solution for the
radial velocity $v(R)$, angular rotation frequency $\Omega(R)$, and
isothermal sound speed $c_s(R)$ of an advection-dominated flow (see
also Spruit et al. 1987 who derived a similar solution in the
context of their discussion of spiral shocks):
$$
v(R) =-{(5+2\epp)\o3\alpha^2}g(\alpha,\epp)\alpha v_{ff}
  \equiv -c_1\alpha v_{ff},
$$
$$
\Omega(R) = \left[{2\epp(5+2\epp)\o9\alpha^2}g(\alpha,\epp)\right]^{1/2}
  {v_{ff}\o R}\equiv c_2{v_{ff}\o R},
$$
$$
c_s^2(R) = {2(5+2\epp)\o9\alpha^2}g(\alpha,\epp)v_{ff}^2
  \equiv c_3v_{ff}^2,
\eqno(2.1)
$$
where
$$
v_{ff} \equiv \left({GM\o R}\right)^{1/2},
$$
$$
\epp \equiv {\ep\o f} = {1\o f}\left({5/3-\gamma\o\gamma-1}\right),
$$
$$
g(\alpha,\epp) \equiv \left[1+{18\alpha^2\o(5+2\epp)^2}\right]^{1/2}-1.
\eqno(2.2)
$$
As in Paper 1, $\alpha$ represents the standard viscosity parameter
(Shakura \& Sunyaev 1973) and $\gamma$ is the ratio of specific heats.
The parameter $f$, which lies in the range 0 to 1, is the fraction of
viscously dissipated energy which is advected; a fraction $1-f$ of the
energy is locally radiated.  The density and pressure in this solution
are given by
$$
\rho ={\dot M\o4\pi RH |v|},\qquad\qquad p = \rho c_s^2,
\eqno(2.3)
$$
where $H$ is the vertical scale height, given approximately by
$$
H/R \approx (2.5c_3)^{1/2}.
\eqno(2.4)
$$
The viscous dissipation of energy per unit volume is
$$
q^+ = {3\epp\rho|v|c_s^2\o2R}.
\eqno(2.5)
$$
The dissipation per unit surface area is $Q^+=2Hq^+$.

We use the above formulae in this paper to calculate the local
properties of the accretion solutions.  For a given choice of
$\alpha$, $\gamma$, and $f$, these relations specify all the local
properties of an advection-dominated flow (except the temperature, for
which we need an equation of state).  We note that, although these
solutions were derived using the height-averaged slim disk equations,
they are in excellent agreement with numerical solutions of a more
exact set of equations which do not make the height-averaging
approximation (Paper 2).  Furthermore, the solutions are valid even
for cooling-dominated flows with $f\rightarrow0$; in this limit the
solution reduces to a reasonable approximation of a thin accretion
disk (Papers 1, 2).

Technically, equations (2.1), (2.2) are valid only for self-similar
flows extending from $R=0$ to $R=\infty$.  In Paper 1 we presented
some numerical solutions of the slim disk equations where we imposed
boundaries at finite radii.  We have also obtained other numerical
solutions where we imposed a sonic boundary condition at the inner
edge, appropriate to a black hole.  In all cases we find that
the numerical solutions quickly settle down to the self-similar form
a short distance away from the boundaries.  This shows that the
self-similar solution is in some sense the ``natural'' state of the
flow.  Another requirement for the validity of the self-similar
solution is that $f$ must be independent of $R$.  But, once again, we
have confirmed with numerical experiments that if $f$ varies (slowly)
with $R$, then the numerical solution tracks the local self-similar
form quite accurately.  As a result of these and other tests we
believe that eqs (2.1) and (2.2) are perfectly adequate for the
calculations presented in this paper.

Finally, we mention that the solutions have been derived using a
Newtonian potential.  For accretion onto black holes and neutron
stars, which is the primary focus of this paper, we should technically
carry out the calculations in the Schwarzschild or Kerr metric.  This
does not appear to be a serious defect and we feel that it will
introduce only small quantitative errors in some of the conclusions.

\bigskip\noindent
{\bf 2.2. Thermodynamics of the Gas}
\bigskip

We assume that the accreting gas is roughly in equipartition with an
isotropically tangled magnetic field, and write the total pressure $p$
as
$$
p=p_g+p_m,\qquad p_g=\beta p,\qquad p_m=(1-\beta)p,
\eqno(2.6)
$$
where $p_g$ is the gas pressure, $p_m$ is the magnetic pressure, and
we take $\beta$ to be independent of $R$.  If we have flux-freezing,
then we know that the magnetic field strength scales with length as
$B\sim\ell^{-2}$.  This means that $p_m\sim B^2\sim\rho^{4/3}$ and the
magnetic field behaves just like radiation.  Assuming such a frozen-in
field, we take $\gamma$ in eq (2.2) to be given by
$$
\gamma=\Gamma_1={32-24\beta-3\beta^2\o24-21\beta},
\eqno(2.7)
$$
where $\Gamma_1$ is the effective $\Gamma$ in the adiabatic $p\rho$
relation as defined by Clayton (1983).  Note that the definition of
$\beta$ in equation (2.6) is similar to that used in discussions of
gas-radiation mixtures (e.g. Clayton 1983), but differs from the usual
definition of $\beta$ in the magnetohydrodynamics literature (the MHD
$\beta$ is equal to our $\beta/(1-\beta)$).  The internal energy per
unit volume of the gas is
$$
U={3\o2}\beta p_g + {B^2\o4\pi}.
\eqno(2.8)
$$

Ever since the important work of SLE, it has been common to discuss
accretion flows in high energy astrophysics in the context of a
two-temperature plasma, where the ions and the electrons have
different temperatures.  In keeping with this practice, we allow the
ion temperature $T_i$ and the electron temperature $T_e$ to be
different, and take the gas pressure of the accreting gas to be given
by
$$
p_g=\beta \rho c_s^2={\rho kT_i\o\mu_im_u}+{\rho kT_e\o\mu_em_u}.
\eqno(2.9)
$$
The effective molecular weights of the ions and electrons are given
respectively by
$$
\mu_i={4\o1+3X}=1.23,\qquad \mu_e={2\o1+X}=1.14,
\eqno(2.10)
$$
where the numerical values correspond to a hydrogen mass fraction
$X=0.75$.  The equipartition magnetic field is determined from the
magnetic pressure:
$$
p_m=(1-\beta)\rho c_s^2={B^2\o8\pi}.
\eqno(2.11)
$$
We do not include radiation pressure in our equation of state because
the optically thin advection-dominated flows we consider always have
gas pressure much larger than radiation pressure.  If we wish to
extend these studies to higher accretion rates, such as those
considered by Abramowicz et al. (1988, 1995), we will need to include
an additional radiation pressure term in equation (2.6).

\bigskip\noindent
{\bf 2.3. Scaled Numerical Relations}
\bigskip

We scale masses in solar units by writing
$$
M=mM_\odot,
\eqno(2.12)
$$
and accretion rates in Eddington units,
$$
\dot M=\md \dot M_{Edd},\qquad
\dot M_{Edd}={L_{Edd}\over\eta_{eff}c^2}=
{4\pi GM\o\eta_{eff}\kappa_{es}c}=1.39\times10^{18}m
{}~{\rm gs}^{-1},\qquad \eta_{eff}=0.1,
\eqno(2.13)
$$
where $\kappa_{es}=0.4{\rm cm^2g^{-1}}$ and we have assumed the
standard value for the accretion efficiency factor $\eta_{eff}$.
Also, we scale radii in Schwarzschild gravitational units,
$$
R=\rs R_{Schw},\qquad R_{Schw}={2GM\o c^2}=2.95\times 10^5m ~{\rm cm}.
\eqno(2.14)
$$
Note that we have introduced the efficiency $\eta_{eff}$ only for easy
comparison with other works. Our accretion flows generally do {\it
not} radiate with an efficiency $\eta_{eff}$.  In fact, in extremely
advection-dominated flows around black holes, the actual radiation
efficiency is very low because the black hole swallows all the
advected energy.

With the above scalings, we can rewrite the equations of \S\S2.1,2.2
as follows:
$$
v  = -2.12\times10^{10}\alpha c_1\rs^{-1/2} ~{\rm cm s}^{-1},
$$
$$
\Omega  = 7.19\times10^4c_2m^{-1}\rs^{-3/2} ~{\rm s}^{-1},
$$
$$
c_s^2  = 4.50\times10^{20}c_3\rs^{-1} ~{\rm cm^2s^{-2}},
$$
$$
\rho  = 3.79\times10^{-5}\alpha^{-1}c_1^{-1}c_3^{-1/2}m^{-1}\md\rs^{-3/2}
  ~{\rm g cm^{-3}},
$$
$$
p  = 1.71\times10^{16}\alpha^{-1}c_1^{-1}c_3^{1/2}m^{-1}\md\rs^{-5/2}
  ~{\rm g cm^{-1} s^{-2}},
$$
$$
B = 6.55\times10^8\alpha^{-1/2}(1-\beta)^{1/2}c_1^{-1/2}c_3^{1/4}
  m^{-1/2}\md^{1/2}\rs^{-5/4} ~{\rm G},
$$
$$
q^+ = 1.84\times10^{21}\epp c_3^{1/2}m^{-2}\md\rs^{-4} ~{\rm erg cm^{-3}
  s^{-1}},
$$
$$
n_e  = \rho/\mu_em_u=2.00\times10^{19}\alpha^{-1}c_1^{-1}c_3^{-1/2}m^{-1}
  \md\rs^{-3/2} ~{\rm cm^{-3}},
$$
$$
\tau_{es}  = 2n_e\sigma_TH = 12.4\alpha^{-1}c_1^{-1}\dot m r^{-1/2}.
\eqno(2.15)
$$
where $n_e$ is the electron number density, $\tau_{es}$ is the
scattering optical depth, $\sigma_T=6.62\times10^{-25} ~{\rm cm^2}$ is
the Thomson cross-section, and the constants $c_1, ~c_2, ~c_3$ are
defined in eq (2.1).
%({\bf It may be useful to give at this point
%actual numerical values of $c_1$, $c_2$, $c_3$ for a particular case,
%e.g. $\beta=0.5$, $f=1$.})
In the case of a two-temperature plasma,
equations (2.9) and (2.10) show that the ion and electron temperatures
satisfy
$$
T_i+1.08T_e = 6.66\times10^{12}\beta c_3\rs^{-1}\ {\rm K}.
\eqno(2.16)
$$

\vfill\eject

\noindent{\bf 3. Heating and Cooling of a Two-Temperature Plasma}
\bigskip\bigskip

We determine the ion and electron temperatures in the accreting plasma
by taking into account the detailed balance of heating, cooling, and
advection.  Due to the large mass difference between ions and
electrons, we expect $q^{+}$ to act primarily on the ions (e.g. SLE,
Rees et al. 1982), which then transfer some of their energy to the
electrons. In this paper, we assume that the transfer from ions to
electrons occurs through Coulomb coupling (SLE) and we ignore other
non-thermal coupling mechanisms which may be present in turbulent
magnetized plasmas (e.g. Begelman \& Chiueh 1988, see Appendix A of
the present paper).  The cooling of the plasma is primarily via
electrons and occurs through a variety of channels as we discuss
below.

\bigskip\noindent
{\bf 3.1. Heating of Electrons by Ions}
\bigskip
If the ions are at a higher temperature than the electrons, Coulomb
collisions transfer energy from ions to electrons at a volume transfer
rate (Stepney \& Guilbert 1983)
$$
q^{ie} = {3\o2}{m_e\o m_p}n_en_i\sigma_Tc
  {(kT_i-kT_e)\o K_2(1/\theta_e)K_2(1/\theta_i)}\ln\Lambda
\qquad\qquad\qquad\qquad\qquad\qquad
$$
$$
\qquad\qquad\qquad\qquad
\times\left[{2(\theta_e+\theta_i)^2+1\o(\theta_e+\theta_i)}
  K_1\left({\theta_e+\theta_i\o\theta_e\theta_i}\right)
  +2K_0\left({\theta_e+\theta_i\o\theta_e\theta_i}\right)\right]
{\rm erg cm^{-3} s^{-1}},
\eqno(3.1)
$$
where the $K$'s are modified Bessel functions, the Coulomb logarithm
is roughly $\ln\Lambda=20$, and the dimensionless electron and ion
temperatures are defined by
$$
\theta_e={kT_e\o m_ec^2},\qquad \theta_i={kT_i\o m_pc^2}.
\eqno(3.2)
$$
Equation (3.1) assumes that all the ions are protons.  For a more
general case, we should replace $n_i$ by a sum over ion species, $\sum
Z_j^2n_j$, where $Z_j$ is the charge of the $j$-th species and $n_j$
is its number density. This gives a factor of $1.25$ for a composition
of 75\% H and 25\% He.  Thus
$$
q^{ie} = 5.61\times10^{-32}{n_e n_i (T_i-T_e)\o K_2(1/\theta_e)K_2(1/\theta_i)}
\qquad\qquad\qquad\qquad\qquad\qquad\qquad\qquad\qquad
$$
$$
\times\left[{2(\theta_e+\theta_i)^2+1\o(\theta_e+\theta_i)}
  K_1\left({\theta_e+\theta_i\o\theta_e\theta_i}\right)
  +2K_0\left({\theta_e+\theta_i\o\theta_e\theta_i}\right)\right]
  {\rm erg cm^{-3} s^{-1}}.
\eqno(3.3)
$$

\bigskip\noindent
{\bf 3.2. Cooling  of Electrons}
\bigskip

Electrons are cooled by many different processes including bremsstrahlung,
synchro-cyclotron, and Compton cooling off soft photons.

\bigskip\noindent
{\bf Bremsstrahlung Cooling}
\bigskip

The electrons cool both by electron-ion and electron-electron
bremsstrahlung. We write the cooling rate per unit volume as
$$
q_{br}^- = q_{ei}^- + q_{ee}^-,
\eqno(3.4)
$$
where the subscripts $ei$ and $ee$ denote the electron-ion and the
electron-electron rates.  Following Stepney and Guilbert (1983), we
adopt
$$
q_{ei}^- = 1.25n_e^2\sigma_Tc\alpha_fm_ec^2F_{ei}(\theta_e)
  = 1.48\times10^{-22}n_e^2F_{ei}(\theta_e) ~{\rm erg cm^{-3} s^{-1}},
\eqno(3.5)
$$
where $\alpha_f$ is the fine structure constant and we have included a
factor of $1.25$ for the same reason as in eq (3.3).  The function
$F_{ei}(\theta_e)$ has the approximate form
$$
\eqalignno{
F_{ei}(\theta_e) & = 4\left({2\theta_e\o\pi^3}\right)^{1/2}
  \left[1+1.781\theta_e^{1.34}\right],\qquad & \theta_e<1, \qquad\qquad \cr
& = {9\theta_e\o2\pi}\left[\ln(1.123\theta_e+0.48)+1.5\right],
  & \theta_e>1.\quad\quad
(3.6)
}
$$
The numerical constant $0.48$ in the second expression is quoted as
$0.42$ by Stepney \& Guilbert (1983) but we have changed the value in
order to have a continuous $F_{ei}(\theta_e)$ across $\theta_e=1$.
For the electron-electron bremsstrahlung, Svensson (1982) gives
expressions in the limits $\theta_e<1$ and $\theta_e>1$.  For
$\theta_e<1$,
$$
\eqalignno{
q_{ee}^- & = n_e^2cr_e^2m_ec^2\alpha_f{20\o9\pi^{1/2}}(44-3\pi^2)
  \theta_e^{3/2}(1+1.1\theta_e+\theta_e^2-1.25\theta_e^{5/2}) \cr
& = 2.56\times10^{-22}n_e^2
  \theta_e^{3/2}(1+1.1\theta_e+\theta_e^2-1.25\theta_e^{5/2})
  ~{\rm erg cm^{-3} s^{-1}},
\qquad\qquad\qquad (3.7)
}
$$
while for $\theta_e>1$,
$$
\eqalignno{
q_{ee}^- & = n_e^2cr_e^2m_ec^2\alpha_f24\theta_e
  (\ln2\eta\theta_e+1.28) \cr
& = 3.40\times10^{-22}n_e^2\theta_e
  (\ln 1.123\theta_e+1.28)
  ~{\rm erg cm^{-3} s^{-1}},
\qquad\qquad\qquad\qquad (3.8)
}
$$
where $r_e=e^2/m_ec^2$ is the classical electron radius and
$\eta=\exp(-\gamma_E)=0.5616$.  For $\theta_e>1$, Svensson (1982)
gives a numerical constant of $5/4$ instead of $1.28$ inside the
parentheses; again we have changed the constant to ensure smoothness
across $\theta_e=1$.

\bigskip\bigskip\noindent
{\bf Synchrotron Cooling}
\bigskip

Cooling via synchrotron radiation is not often considered in thermal
models of X-ray binaries (e.g. SLE and Wandel \& Liang 1991 ignore it
but Kusunose \& Takahara 1989 do include it), though it is fairly
common in discussions of AGN models (e.g. the ion torus model of Rees
et al. 1982).  Due to the assumption of an equipartition magnetic
field in the plasma, synchrotron emission from thermal electrons can
be quite important in our model, and we therefore include it in the
calculations.

In the optically thin limit, the spectrum of synchrotron emission by a
relativistic Maxwellian distribution of electrons is given by
Pacholczyk (1970),
$$
\ep_{synch} d\nu = 4.43\times 10^{-30}{n_e\nu\o K_2(1/\theta_e)}
  I\left({x_M\o\sin\theta}\right)d\nu
  ~{\rm erg cm^{-3} s^{-1} Hz^{-1}},
\eqno(3.9)
$$
where $I(x)$ is a tabulated function,
$$
x_M\equiv {2\nu\o3\nu_0\theta_e^2},\qquad
  \nu_0\equiv {eB\o2\pi m_ec},
\eqno(3.10)
$$
and $\theta$ is the angle between the velocity vector of the electron
and the direction of the local magnetic field.  Mahadevan, Narayan \&
Yi (1995) have obtained an extremely accurate fitting function
(maximum error 0.0036) for $I(x_M)$,
$$
I(x_M) = 2.5651\left(1+{1.92\o x_M^{1/3}}+{0.9977\o x_M^{2/3}}
  \right)\exp\left(-1.8899x_M^{1/3}\right).
\eqno(3.11)
$$
Averaging over $\theta$ for an isotropic velocity distribution,
$I(x_M/\sin\theta)$ gets replaced by a new function $I'(x_M)$, for
which the fitting function (max error 0.015) is
$$
I'(x_M) = {4.0505\o x_M^{1/6}}\left(1+{0.40\o x_M^{1/4}}+{0.5316\o x_M^{1/2}}
  \right)\exp\left(-1.8899x_M^{1/3}\right).
\eqno(3.12)
$$
Although these formulae are valid only for relativistic temperatures,
$\theta_e\ \sgreat\ 1$, we use them at all temperatures.  In our
models, the synchrotron emission always comes from relativistic
electrons in the tail of the Maxwellian distribution, even if the
temperature is somewhat sub-relativistic, and so this approximation is
reasonable.

Optically thin synchrotron emission rises steeply with decreasing
frequency $\nu$. Under most circumstances, the emission is
self-absorbed below a critical frequency $\nu_c$.  We estimate $\nu_c$
by the following approximate calculation.  At radius R, we assume that
the synchrotron emission occurs over the volume of a sphere of radius
R, and we equate this emission to the Rayleigh-Jeans blackbody
emission from the surface of the sphere.  This gives the condition
$$
4.43\times10^{-30}{n_e\nu\o K_2(1/\theta_e)}I'(x_M)\cdot {4\pi\o 3} R^3 =
  \pi\cdot2{\nu^2\o c^2}kT_e\cdot 4\pi R^2.
\eqno(3.13)
$$
Substituting for $I^{\prime}(x_M)$, we then obtain a transcendental
equation for $x_M$,
$$
\exp\left(1.8899x_M^{1/3}\right) = 2.49\times10^{-10}{n_eR\o B}
  {1\o\theta_e^3K_2(1/\theta_e)}\left({1\o x_M^{7/6}}
  +{0.40\o x_M^{17/12}}+{0.5316\o x_M^{5/3}}\right).
\eqno(3.14)
$$
We solve this equation numerically at each $R$ to obtain $x_M$,
and thereby calculate the critical frequency $\nu_c$ by means of the
relation
$$
\nu_c = {3\o2}\nu_0\theta_e^2x_M,\qquad \nu_0=2.80\times10^6B
{\rm Hz}, \eqno(3.15)
$$
where $B$ is expressed in Gauss. Technically, the above calculation is
valid only for a spherical flow and we need to introduce corrections
of order $H/R$ for a rotationally flattened flow. However, synchrotron
emission is important only in the hot highly advection-dominated
solutions and these are essentially spherical (Paper 2).  We therefore
feel these formulae are adequate.

To estimate the cooling per unit volume due to synchrotron radiation
we use the relation
$$
q_{synch}^-\cdot4\pi R^2\Delta R = \pi\cdot2{\nu_c^2(R)\o c^2}kT_e(R)
  {d\nu_c(R)\o dR}\Delta R\cdot 4\pi R^2.
\eqno(3.16)
$$
The left hand side of this equation is the total cooling over a shell
extending from radius $R$ to $R+dR$.  The right-hand side represents
the net flux reaching an observer at infinity from this shell.  This
expression is obtained by assuming that at each frequency $\nu$ the
observer sees a blackbody source with a radius determined by the
condition $\nu=\nu_c(R)$.  We thus obtain the cooling rate per unit
volume to be
$$
q_{synch}^-={2\pi\o 3 c^2}kT_e(R){d\nu_c^3(R)\o dR}.
\eqno(3.17)
$$
This result is valid only if $\nu_c$ increases monotonically with
decreasing $R$, which is satisfied by all our models.  Equation (3.17)
is a non-local expression for the synchrotron emissivity since it
depends on properties at neighboring $R$. For even greater simplicity,
we often adopt a purely ``local'' approximation, in which the
derivative of $\nu_c^3$ with respect to $R$ is simply replaced by
$\nu_c^3/R$. Thus we have approximately
$$
q_{synch}^-\approx{2\pi\o 3 c^2}kT_e(R){\nu_c^3(R)\o R}.
\eqno(3.18)
$$

Note that the simple treatment of synchrotron emission discussed here
ignores energy transport from one radius to another by radiative
diffusion of the self-absorbed photons.  It is not clear how large
this effect is.  Note also that the cooling rate given here is
different from other expressions given in the literature (e.g. Ipser
\& Price 1983).  Equation (3.17) has been written in a form that
explicitly ensures that the integral of $q_{synch}^-$ over the entire
flow is equal to the total cooling radiation that reaches infinity.

\bigskip\noindent
{\bf Compton Cooling}
\bigskip

Although we neglect nonlocal radiative transfer effects in the
optically thin situations considered in this paper, we cannot afford
to ignore the scattering of escaping photons off hot thermal
electrons.  Indeed, in some of our models the scattering depth is
comparable to unity and Comptonization of soft photons by hot
electrons becomes an important cooling mechanism, especially in the
inner regions of the flow.

Dermer, Liang \& Canfield (1991) have given a convenient though
approximate prescription for the Comptonized energy enhancement factor
$\eta$, which is defined to be the average change in energy of a
photon between injection and escape. Their prescription is
$$
\eta = 1+{P(A-1)\o(1-PA)}\left[1-\left({x\o3\theta_e}\right)
  ^{-1-\ln P/\ln A}\right]\equiv
  1+\eta_1-\eta_2\left({x\o\theta_e}\right)^{\eta_3},
\eqno(3.19)
$$
where
$$
\eqalignno{
x & = h\nu/m_ec^2, \cr
P & = 1-\exp(-\tau_{es}), \cr
A & = 1 + 4\theta_e + 16\theta_e^2.
\qquad\qquad\qquad\qquad\qquad (3.20)
}
$$
The factor $P$ is the probability that an escaping photon is
scattered, while $A$ is the mean amplification factor in the energy of
a scattered photon when the scattering electrons have a Maxwellian
velocity distribution of temperature $\theta_e$.  Note that the extra
cooling due to Comptonization is $\eta-1$ times the mean flux of
escaping photons.  The formula (3.19) is valid only for soft photons,
i.e. for $x<3\theta_e$.  Hard photons actually heat up the electrons,
but we ignore this since our models generally have very little hard
radiation capable of causing Compton-heating.

An important feature of the Dermer et al. formula is that it treats
both the unsaturated and saturated limits of Comptonization. SLE
considered only the unsaturated Comptonization limit where the Compton
$y$-parameter is taken to be around unity.  They also assumed an
external supply of soft photons and solved the steady-state Kompaneets
equation to obtain the spectrum (cf.  Eardley et al. 1978).  The
formulae given above use the Thomson cross-section, whereas one should
technically employ the Klein-Nishina cross-section at these
temperatures.  In practice, this simplification is not serious.  The
bulk of the Comptonization occurs on very soft photons, either
synchrotron radiation or thermal radiation from the accreting star,
and these photons remain soft even in the rest frame of the electrons.
Therefore, for single or even double Compton scattering the Thomson
approximation is very good.  Saturated Comptonization occurs only for
limiting solutions, near the critical $\dot m_{crit}$ lines discussed
in \S4; in these solutions the electron temperature tends to be a
little lower, $T_e\ \sles\ 10^{9.5}$ K, where the Klein-Nishina
correction is less important.

In our models, Comptonization can operate on three different sources
of radiation, and we need to calculate the cooling due to each of
these.

\medskip\noindent
{\it Comptonization of Bremsstrahlung Radiation}: We have earlier
estimated the total cooling $q_{br}^-$ due to bremsstrahlung.  The
spectrum of the emission per unit volume may be written as a function
of $x=h\nu/m_ec^2$ as follows:
$$
\ep_{br}\left({x\o\theta_e}\right)d\left({x\o\theta_e}\right)=
  q_{br}^-\exp\left(-{x\o\theta_e}\right)d\left({x\o\theta_e}\right).
\eqno(3.21)
$$
This spectrum, which corresponds to the electron-ion emission, may not
adequately describe electron-electron bremsstrahlung, but the latter
is not important at the temperatures we are interested in.  For
computational convenience, we follow Dermer et al. (1991) and replace
eq (3.20a) by the following simple function,
$$
\eqalignno{
\epsilon_{br}\left(x\o \theta_e\right)d\left(x\o \theta_e\right) & =
q^-_{br}d\left(x\o \theta_e\right),\qquad {x\o \theta_e}\le 1 \cr
& =0,\qquad\qquad\qquad {x\o \theta_e}>1.
\qquad\qquad\qquad\qquad\qquad\qquad\qquad\qquad
(3.22)}
$$
The part of the spectrum which can be Comptonized is the range from
$x=x_c=h\nu_c/ m_ec^2$ at the synchrotron self-absorption edge upto
$x=\theta_e$.  Thus, the volume rate of cooling of electrons by
self-Comptonization of bremsstrahlung radiation is
$$
q_{br,C}^- = 3\eta_1 q_{br}^-\left[
\left({1\o 3}-{x_c\o3\theta_e}\right)-{1\o \eta_3+1}
\left(\left(1\over 3\right)^{\eta_3+1}-\left(1\o
3\theta_e\right)^{\eta_3+1}\right)\right].
\eqno(3.23)
$$

\medskip\noindent
{\it Comptonization of Synchrotron Radiation}: In our simple model of
synchrotron cooling, the escaping radiation is emitted mostly at the
local self-absorption cut-off frequency $\nu_c$ given by equation
(3.15).  The Comptonization of this radiation gives an additional
cooling rate
$$
q_{synch,C}^- = q_{synch}^-\left[\eta_1-\eta_2\left({x_c\o\theta_e}
  \right)^{\eta_3}\right].
\eqno(3.24)
$$

\medskip\noindent
{\it Comptonization of Soft photons from the Star}: A third source of
soft photons is radiation emitted by the accreting star.  SLE were the
first to note that Comptonization of externally supplied soft photons
can be a very efficient cooling process.  The soft photons may arise
from a standard thin accretion disk, or from the companion star in a
binary.  We consider here soft photons which may be emitted by the
accreting star.  In principle, the star may have its own internal
source of luminosity.  However, we assume that the dominant stellar
luminosity is that which arises from the accretion itself, viz. the
internal energy of the hot accreting material which is thermalized and
radiated at the stellar surface.  The luminosity due to this component
is
$$
L_* = 4\pi R_*H(R_*)|v(R_*)|\left[U(R_*)+{1\o2}\rho(R_*)v^2(R_*)\right],
\eqno(3.25)
$$
where $R_*$ is the radius of the star and $U$ is the internal energy
of the accreting gas (eq 2.11).  In eq (3.25) we have included the
radial kinetic energy of the gas in calculating $L_*$ but not the
rotational energy.  This is because we assume that the star is
spinning in equilibrium with the accreting material.

It is unclear at this time exactly what the spectrum of the re-emitted
radiation will be.  One possibility is that all the incoming internal
energy in the accretion flow is fully thermalized at the stellar
surface so that the star radiates essentially as a blackbody.  In that
case, the temperature of the radiation from the star is given by
$$
T_* = \left({L_*\o4\pi R_*^2\sigma}\right)^{1/4}.
\eqno(3.26)
$$
However, it is quite possible that thermalization is incomplete and
that the radiation comes out with a much harder spectrum than a
blackbody at $T_*$.  The outgoing flux at radius $R$ is given by
$$
F_*(R)={L_*\o4\pi R^2}.
\eqno(3.27)
$$
If the stellar radiation is blackbody, then in the same spirit as eq
(3.22), we can approximate the spectrum as a pure Rayleigh-Jeans
$\nu^2$ spectrum with a sharp cut-off at $\nu_b$, where $\nu_b$ is
given by
$$
\sigma T_*^4=\pi\int_0^{\nu_b}{2{\nu^2\o c^2}kT_*d\nu}
\quad\rightarrow\quad \nu_b=5.61\times10^{10}T_* ~{\rm s^{-1}}.
\eqno(3.28)
$$
We thus write the spectrum approximately as
$$
\eqalignno{
F\left({x\o\theta_e}\right)d\left({x\o\theta_e}\right) & =
  3F_*\left({\theta_e\o x_b}\right)^3\left({x\o\theta_e}\right)^2
  d\left({x\o\theta_e}\right),\qquad x\leq x_b=  {h\nu_b\o m_ec^2},\cr
& = 0,\qquad\qquad\qquad\qquad\qquad\qquad\qquad x>x_b.
  \qquad\qquad\qquad (3.29)}
$$
The part of the flux between $x=x_c$ and $x=x_b$ will be Comptonized
provided $x_b<3\theta_e$ (Wien limit).  If $x_b>3\theta_e$, then only
the part from $x=x_c$ to $x=3\theta_e$ is Comptonized. We ignore
Compton heating.

We can calculate the Comptonization of the stellar flux just as we did
the brems-strahlung case, except that we must replace $\ep_{br}$ by
$F/R$.  Thus, we obtain a cooling rate
$$
q_{*,C}^- = 3{F_*\o R}\left({\theta_e\o x_b}\right)^3
  \left[{\eta_1\o3}\left\{\left({x_{max}\o\theta_e}\right)^3-
  \left({x_c\o\theta_e}\right)^3\right\}-{\eta_2\o3+\eta_3}
  \left\{\left({x_{max}\o\theta_e}\right)^{3+\eta_3}-
  \left({x_c\o\theta_e}\right)^{3+\eta_3}\right\}\right],
\eqno(3.30)
$$
where $x_{max}={\rm max}(x_b,3\theta_e)$.  Note that the cooling term
$q_{*,C}^-$ is included only in the case of accretion on a normal star
such as a neutron star.  We do not include this term for accretion on
a black hole which lacks a radiating surface.  Note also that the
analysis we have presented is valid only when the accreting gas is
optically very thin.  In some of the solutions described in this paper
the optical depth approaches unity and we expect substantial radiative
transfer effects in the propagation of the soft photons from the star.
It is beyond the scope of this paper to model these effects accurately.

\bigskip\noindent
{\bf Optically Thick Cooling}
\bigskip

For the advection-dominated flows that we concentrate on in this
paper, it is sufficient to set the volume rate of cooling equal to the
sum of the five cooling rates given above, viz. $q_{br}^-$,
$q_{synch}^-$, $q_{br,C}^-$, $q_{synch,C}^-$, $q_{*,C}^-$.  However,
we also calculate solutions corresponding to normal thin accretion
disks.  These solutions are optically thick and radiate like a
blackbody or a modified blackbody.  Since this is a very different
regime, we need to generalize our cooling formula.

Hubeny (1990) has shown that the effective surface flux $F_v$ from
an accretion disk can be written approximately as
$$
F_v=\sigma T_{eff}^4={4\sigma T_e^4\o
{3\tau\o 2}+\sqrt{3}+{1\over\tau_{abs}}},
\eqno(3.31)
$$
where $\tau=\tau_{abs}+\tau_{es}$ is the total optical depth in the
vertical direction from the disk midplane to the surface, $\tau_{abs}$
is the corresponding absorption optical depth, $T_{eff}$ is the
effective surface temperature, and $T_{e}$ is the electron temperature
at the equatorial plane.  Equation (2.15) gives an expression for
$\tau_{es}$, but to calculate $\tau_{abs}$ we need to evaluate all the
relevant emission and absorption processes.  Note however that, in the
extreme optically thin limit, eq (3.31) gives $F_v=4\sigma
T_e^4\tau_{abs}$.  Since the previous sections give the net cooling in
precisely this limit, we can therefore estimate $\tau_{abs}$
approximately by
$$
\tau_{abs}={H\over 4\sigma
T_{e}^4}\left(q_{br}^-+q_{synch}^-+q_{br,C}^-+q_{synch,C}^-+q_{*,C}^-
\right).
\eqno(3.32)
$$
Substituting this estimate of $\tau_{abs}$ back into eq (3.31), we
obtain a modified expression for the net volume cooling rate of the
accreting gas:
$$
q^-={F_v\o H}={4\sigma T_e^4/H\o {3\tau\o 2}+\sqrt{3}+
{4\sigma T_e^4\o
H}(q^-_{br}+q^-_{synch}+q^-_{br,C}+q^-_{synch,C}+q^-_{*,C})^{-1}
}.
\eqno(3.33)
$$
Equation (3.33) is valid both in the optically thick and thin limits.
When the gas is extremely optically thick, this formula gives
$q^-=8\sigma T_e^4/3H\tau$ which is the appropriate blackbody limit,
whereas in the optically thin limit it gives
$q^-=q^-_{br}+q^-_{synch}+q^-_{br,C}+q^-_{synch,C}+q^-_{*,C}$. The
formula thus provides a convenient interpolation between the two
limits.

\bigskip\noindent
{\bf 3.3. Thermal Balance Equations}
\bigskip

Using the heating and cooling rates given above, we are in a position
to solve for the energy balance of the accreting medium.  In our
model, a fraction $f$ of the viscously dissipated energy $q^+$ is
advected inward while a fraction $1-f$ is transferred from ions to
electrons and then radiated.  We therefore have two relations which
finally close the set of equations and help determine the ion and
electron temperatures.

Consider first the energy balance of the ions.  We require the rate of
input of energy through viscous dissipation to be equal to the sum of
the rate of advection of energy by the ions and the rate of transfer
of energy from the ions to the electrons.  This gives the condition
$$
q^+ = q^{adv} + q^{ie} = fq^+ + q^{ie},
\eqno(3.34)
$$
where $q^+$ and $q^{ie}$ are given in eqs (2.5) and (3.3), and we
set $q^{adv}=fq^+$ by the definition of $f$.  In writing this
equation we have assumed that all the viscous energy goes into the
ions, which is reasonable since the ions are very much more massive
than the electrons.

We obtain a second equation by considering the energy balance of the
electrons.  In this case we require the net heating and cooling rates
of the electrons to be equal, which gives
$$
q^{ie}=q^-,
\eqno(3.35)
$$
where $q^-$ is calculated via eq (3.33).

For a given $M$, $\dot M$, $R$, $\alpha$, $\beta$, we can think of
equations (2.16), (3.34) and (3.35) as three relations which may be
solved to obtain the ion and electron temperatures, $T_i$, $T_e$, and
the advective fraction $f$.  The other parameters of the gas such as
$v$, $\rho$, $B$, etc., which are needed to calculate $q^+$, $q^{ie}$
and $q^-$, are given in eq (2.15).  Since the equations are fairly
complex we solve them numerically.  The important point however is
that the solutions we obtain are fully self-consistent within the
framework of our model.  Indeed, considering that we have included
advection, and that we model a two-temperature plasma along with a
variety of cooling mechanisms including synchrotron emission and
Comptonization with proper treatment of saturation, the calculations
presented here may well be the most complete and self-consistent
computations attempted of a high energy accretion flow.

A point worth emphasizing is the presence of the advection term
$q^{adv}$ in equation (3.34).  It is this term above all else,
coupled with the use of the self-similar solution discussed in
\S2, which sets this work apart from earlier analyses of accretion
flows.  In virtually all previous studies it has been the practice to
assume local energy balance.  This means that in single temperature
models one requires $q^+=q^-$, while in two-temperature models one
tries to satisfy $q^+=q^{ie}=q^-$.  Our solutions are more general
because they allow for advection.  Moreover, the fraction $f$ of the
energy which is advected is not assigned arbitrarily but is solved for
self-consistently, as are the ion and electron temperatures and other
properties of the accreting gas.  Because of this we are able to
reproduce the two previously known solutions, namely the
Shakura-Sunyaev thin disk and the SLE hot disk, both of which have
negligible advection ($f\ll1$), and in addition we obtain a new class
of very hot solutions which are advection-dominated ($f\rightarrow1$).

\vfill\eject
\noindent{\bf 4. Results}
\bigskip\bigskip

As we have discussed in \S1 (see also Papers 1, 2),
advection-dominated accretion can arise both in the optically thin and
optically thick limits.  In this paper we restrict our investigation
to the optically thin case.  The paper by Rees et al. (1982) on ion
tori showed that there is an upper limit to the mass accretion rate
such that only below this limit is optically thin advection-dominated
accretion possible.  Above the limiting $\dot M$, the cooling is too
efficient, and the only configuration allowed for the flow is the
standard thin accretion disk.  We explore this point here
quantitatively using the detailed formalism described in the previous
two sections.

\bigskip
\noindent{\bf 4.1. Critical Accretion Rate for a Special Case}
\bigskip

We begin by discussing a particularly simple case where the accreting
material is cool and single-temperature, and the dominant cooling
process is bremsstrahlung.  This case has been considered by
Abramowicz et al. (1995) and Chen (1995) who derived a formula for the
maximum $\dot M$ up to which an advection-dominated solution is
possible.  To derive the formula, we choose a value for the parameter
$f$, the fraction of the dissipated energy which is advected with the
flow.  We then equate the local bremsstrahlung cooling rate,
$q^{-}_{br}$, to a fraction $1-f$ of the heating rate $q^{+}$ (see
equation 3.34).  Setting $T_i=T_e$ in eq (2.16) and using the
non-relativistic limit of eqs. (3.6) and (3.7), this calculation gives
for the scaled accretion rate (see eqs 2.12--2.14 for the scalings)
$$
{\dot m}=1.32\times 10^3(1-f)\epp\alpha^2\beta^{-1/2}c_1^2c_3
r^{-1/2}.\eqno(4.1)
$$
Let us set $f=1/2$ in this equation.  This corresponds to half the
released energy being advected and half being radiated, which would
appear to be a reasonable choice for this calculation.  Assuming
$\alpha=0.3$, $\beta=0.5$ as a specific example, eq (4.1) gives
${\dot m}=4.5r^{-1/2}$ for the limiting accretion rate up to which an
optically thin advection-dominated solution is possible.  This result
is roughly similar to that obtained by Abramowicz et al.  (1995) and
Chen (1995).  We see that the critical $\dot m$ when expressed in
scaled units does not depend explicitly on the mass of the accreting
star.

We can compare the analytical result (4.1) with more elaborate
numerical results obtained by solving the equations written down in
\S\S2,3.  Figure 1 shows the numerically calculated $\dot m$ as a
function of scaled radius $r$ for a $10M_\odot$ black hole.  The
calculation assumes $\alpha=0.3$, $\beta=0.5$, $f=0.5$ as before and
the exact results are compared with the analytical formula (4.1).  We
see that the formula (4.1) is in excellent agreement with the
numerical results at large radii, $r\ge 10^3$, where the plasma is
single-temperature and the cooling is indeed dominated by
bremsstrahlung as assumed.  However, there are significant differences
for $1\le r\le 10^3$.  This is because the plasma here is much hotter,
the ion-electron coupling is weaker, and cooling processes are much
more complex.  Therefore, it is necessary to allow for a
two-temperature plasma and to keep all terms in the energy balance
equations.

\bigskip
\noindent{\bf 4.2. Three Solution Branches}
\bigskip

Before we proceed further, we briefly discuss the relation between
the advection-dominated flows which we focus on in this paper and the
standard cooling-dominated flows which are generally discussed in the
literature.  Formally, in our models, the parameter $f$ distinguishes
between the two types of flow; advection-dominated flows have
$f\sim1$ while cooling-dominated flows have $f\ll1$.  We show now
that, under a given set of conditions, one or the other, or possibly
both, of these kinds of solutions are allowed for an accretion flow.

As an example, we consider the case of an accreting $10M_\odot$ black
hole with $\alpha=0.3$, $\beta=0.5$, and focus on some particular
radius, say $r=10^3$.  Given these parameters, our model allows us to
solve for $f$ as a function of $\dot m$, or vice versa; that is, the
equations written down in \S\S2,3 define a unique mapping between $f$
and $\dot m$.  We show in Fig. 2 the nature of the mapping for the
case under consideration.  Both at very low $\dot m$ and very high
$\dot m$ there is only one solution allowed for $f$ and therefore
there is only one kind of flow allowed.  However, for intermediate
values of $\dot m$ we see that there are {\it three} separate
solutions.  The uppermost branch, with a large value of $f$, is the
advection-dominated branch which we study in this paper, and which has
been discovered independently by Abramowicz et al. (1995).  The
lowermost branch has a very low value of $f$ and is therefore
dominated by cooling.  This solution has a low temperature, is
optically thick, and corresponds to the well-known standard thin
accretion disk solution (e.g. Shakura \& Sunyaev 1973, Frank et
al. 1992).  Both of these branches are stable at this radius.  In
addition, there is a middle branch, which is thermally unstable (see
\S4.7), and which we indicate by a dotted line.  This solution has a
low value of $f\ll1$ and is therefore cooling-dominated (though its
value of $f$ is larger than that of the Shakura-Sunyaev solution).
This solution is much hotter than the thin disk branch and is, in
fact, the unstable SLE hot solution.

The upper branch in Fig. 2 does not extend above a certain maximum
accretion rate $\dot m_{crit}$.  In this particular example
corresponding to $r=10^3$, the critical configuration corresponds to
$\dot m_{crit}=0.16$, $f=0.27$.  Note that in the simplified
discussion of \S4.1 we arbitrarily set $f=0.5$ in order to derive eq
(4.1) and also to calculate the numerical line shown in Fig. 1.  The
correct procedure is to vary $f$ and to optimize it so as to determine
the true maximum or critical accretion rate.  We employ this procedure
to derive all the results presented in the sections below.

We see from Fig. 2 that there is a second critical accretion rate, a
minimum $\dot m$ below which the standard cooling-dominated branch
ceases to exist.  This limit, which we designate $\dot m_{crit}^{'}$,
corresponds to the point where the optical depth $\tau$ of a thin
accretion disk becomes equal to unity
$$
\dot m = 2.6\times10^{-8}\alpha^4\approx\md_{crit}^{'}.\eqno (4.2)
$$
Below this accretion rate, the optical depth falls below unity and the
solution goes unstable by the classic thermal instability associated
with optically thin bremsstrahlung cooling (Pringle et al. 1973, Piran
1978).  The second critical rate $\dot m_{crit}^{'}$ corresponds to
this instability.  Setting $\alpha=0.3$ eq (4.2) gives
$\md\sim10^{-10}$.  The actual limiting $\md$ where the cooling branch
terminates in Fig. 2 is a couple of orders lower because of
differences in the numerical factors.  We should warn the reader that
the opacities we have used are not reliable at the low temperatures at
which this limit occurs, so that the results are not accurate.
However, the existence of the second critical $\dot m_{crit}^{'}$ is
valid regardless of the details of the opacities, so that the overall
topology of Fig. 2 is correct.

\bigskip
\noindent{\bf 4.3. Stellar Mass Black Holes}
\bigskip

Since a black hole lacks a hard radiating surface, energy which is
advected with the accretion flow is lost into the black hole without
being reprocessed and radiated.  This is very different from the case
of a neutron star where all the advected energy must ultimately be
radiated from the stellar surface.  This difference could potentially
lead to observable signatures which may distinguish black holes from
neutron stars (cf. Sunyaev \& Titarchuk 1980, Sunyaev et al. 1991ab,
Liang 1993).  In this and the next subsection we discuss accreting
black holes, and in \S4.5 we discuss accreting neutron stars.

Figure 3 shows some results corresponding to advection-dominated
accretion on a $10M_\odot$ black hole with $\beta=0.5$, where we
remind the reader that the fraction of the total pressure due to
magnetic fields is $(1-\beta)$.  The solid lines in the upper panels
indicate the critical accretion rate $\dot m_{crit}$ as a function of
the radius $r$ for three choices of $\alpha$: 0.03, 0.1, 0.3.  The
procedure we employ to calculate the critical lines is to optimize at
each $r$ the value of $f$ so as to find the maximum $\dot m$ for which
the advection-dominated branch exists at that $r$.  The typical value
of $f$ on the lines is $\sim 0.3$.  We see from Fig. 3 that $\dot
m_{crit}$ depends sensitively on $\alpha$.  For a large $\alpha=0.3$,
for instance, advection-dominated accretion occurs for $\dot m$ as
high as ${\dot m}\sim 0.1$ while for $\alpha=0.03$ we need
$\md<0.001$.  The values of $\dot m_{crit}$ scale essentially as
$\alpha^2$, as shown by equation (4.1) (see also Rees et al. 1982).
In contrast to the $\alpha$ dependence, we find that the solutions are
fairly insensitive to the value of $\beta$.  For instance, if we
change $\beta$ from 0.5 (gas pressure = magnetic pressure) to 0.9 (gas
pressure = 90\% of total pressure), $\md_{crit}$ moves by only
$\sim10\%$.  To indicate how the advection-dominated solutions evolve
as we move down from the critical line we show in Fig. 3 a second set
of lines corresponding to $f=0.9$.  We see that the degree of
advection-domination increases quite rapidly with decreasing $\md$.
Indeed, for $\md\ll\md_{crit}$, the solutions have $f$ practically
equal to unity so that there is virtually no cooling at all.

The lower panels in Fig. 3 show the ion and electron temperatures as a
function of $r$ as we move along the lines corresponding to $\dot
m=\dot m_{crit}$ and $f=0.9$ in the upper panels.  We see that the
flow has two very distinct zones.  For $r\ \sgreat\ 10^3$, the ions
and electrons have almost the same temperature, both being nearly at
the virial temperature.  This is the region where the ions and
electrons are well coupled and bremsstrahlung cooling dominates, so
that the simple expression given in eq (4.1) is valid (see Fig. 1, and
also Abramowicz et al. 1995, Chen 1995).  However, for $r<10^3$, the
temperature becomes very high and the ion-electron coupling becomes
weak.  Other cooling processes also enter the picture.  Synchrotron
emission from thermal electrons dominates over bremsstrahlung, and
also Comptonization of the synchrotron and bremsstrahlung photons
becomes important.  As a consequence of these new effects the plasma
switches to a two-temperature state.  The ion temperature continues to
track the virial temperature and rises monotonically inwards.
However, the electron temperature saturates at a ${\rm
few}\times10^9$K and remains nearly constant all the way down to the
Schwarzschild radius.  The $f=0.9$ curves show that with decreasing
$\dot m$, i.e. increasing $f$, the electron temperature rises in the
inner region.  This is because the density decreases and some of the
cooling processes are less efficient.  Indeed, for highly
advection-dominated flows with $\dot m\ll\dot m_{crit}$, the electron
temperature exceeds $10^{10}$K in our models.  We must caution that
some of our results near the Schwarzschild radius are likely to be
inaccurate, both because we use a Newtonian potential and because we
do not include the expected sonic transition close to the hole.  It
would be of interest to include these effects self-consistently in
future calculations.

\bigskip
\noindent{\bf 4.4. Massive Black Holes}
\bigskip

In Figure 4, we compare the critical accretion rates and the ion and
electron temperatures for black holes of two very different masses,
viz. $10M_{\sun}$ and $10^8M_{\sun}$.  For the same model parameters
($\alpha=0.3$, $\beta=0.5$), the two cases show very similar behavior.
This means that advection-dominated flows are nearly scale-free so
that when the physical variables are scaled in terms of the relevant
fiducial quantities (Eddington accretion rate, Schwarzschild radius)
quantities like the critical accretion rate and the ion and electron
temperatures are essentially independent of the mass.  This is despite
quite large differences in actual physical conditions like density,
magnetic field strength, etc.  The near invariance with respect to
black hole mass suggests that some of the observable emission, such as
bremsstrahlung radiation, from stellar mass black holes and
supermassive black holes may be very similar, whenever these systems
accrete in an advection-dominated mode.  Not all features are similar,
however.  The synchrotron emission, for instance, is radiated at the
critical frequency $\nu_c$ (equation 3.15), which is proportional to
the magnetic field strength $B$ and therefore varies as $m^{-1/2}$
(see equation 2.15).  This gives a fairly strong mass dependence.
Note that the Shakura-Sunyaev thin disk solution has a characteristic
blackbody temperature which scales as $m^{-1/4}$.  This
dependence is mid-way between those shown by the bremsstrahlung and
synchrotron emission in the advection-dominated solution.

\bigskip
\noindent{\bf 4.5. Neutron Stars}
\bigskip

In Figure 5 we show our results for advection-dominated accretion on a
neutron star.  Apart from changing the mass of the accreting star to
$1.4M_\odot$ and the inner radius of the flow to $r_{in} =r_*=2.5$
(corresponding to $R_*=10.5$ km), a crucial difference in this case is
that we allow the star to have a hard surface.  All the advected
energy is assumed to be re-radiated at this surface.  This radiation
moves back through the hot accreting gas, and provides an additional
source of photons for Compton cooling.  In panels {\it a} and {\it b}
of Fig. 5 we assume that the star radiates as a blackbody at
temperature $T_*$ as given in equation (3.26), while in panel {\it c}
we consider for contrast a case where the radiation is emitted at
essentially the electron temperature of the incoming gas, $T_e=10^9$
K.

Black hole candidates among X-ray binaries in general have harder
X-ray spectra than neutron star binaries.  In recent times it has
become accepted that the difference could be the result of extra
cooling due to Comptonization of stellar photons in the neutron star
systems (e.g. Sunyaev et al. 1991ab, Liang 1993, see also Sunyaev \&
Titarchuk 1980).  The results shown in Fig. 5 are an attempt to check
this idea quantitatively, something that apparently has not been done
so far.

Comparing Figs. 3 and 5 we see that there is indeed a substantial
difference between the black hole and neutron star cases, especially
in panels {\it a} and {\it b} where the stellar radiation is quite
cool.  We find that advection-domination is limited to much lower
accretion rates in neutron stars compared to black holes.  The
difference arises primarily because of the very efficient Compton
cooling by the stellar soft photons so that at a given relative accretion
rate, a neutron star system cools more efficiently than a black hole
system.  It is therefore necessary to go to a lower $\md$ before the
system can be advection-dominated.  One consequence of the enhanced
cooling is that, even though a black hole may be more massive (e.g.
$10M_\odot$ vs $1.4M_\odot$), still for a given scaled accretion rate
$\dot m$, neutron star and black hole systems may exhibit similar
luminosities.

The other difference we notice between Figs. 3 and 5 is in the
electron temperature.  In general, $T_e$ is lower in the neutron star
case than in the black hole case.  Once again, cooling by soft photons
is the reason.  This difference may be expected to translate into
fairly clear differences in the spectra, with accreting black holes
likely to have harder spectra.  This statement refers to the emission
from the accretion flow.  In addition, of course, the spectrum of an
accreting neutron star will include unscattered soft thermal radiation
from the stellar surface, whereas this component will be missing in
the black hole case.  One other consequence of the higher electron
temperature in the black hole case is that it may lead to stronger
pair effects at high $\dot m$, and this could produce significant
differences between the two cases.  However, pair effects are not
important at accretion rates $\md<\md_{crit}$.  We have calculated
equilibrium pair densities in our solutions using the equations given
in White \& Lightman (1989, see also Svensson 1984, Zdziarski 1985,
Kusunose \& Takahara 1989), and in no case do we find a positron to
ion ratio $n_+/n_i$ in excess of $\sim 10^{-3}$.

Panel {\it c} in Fig. 5 corresponds to a case where there is the same
amount of stellar radiation as in the previous two panels except that
the radiation is at a temperature of $10^9$ K rather than the
thermalization temperature of $\sim 10^7$ K.  We see that the critical
lines move up a little and the electron temperature in the accretion
flow is also a little higher.  However, the results still do not quite
match the corresponding results for a black hole.  The point is that,
even at such a high temperature, there is still significant
Compton-cooling of the electrons by the stellar radiation.  Therefore,
we see that the effect of the stellar radiation is quite robust, and
fairly insensitive to the temperature of the radiation.

Note that the critical $\dot m_{crit}$ lines in Fig. 5 have a peculiar
jump at $r\sim 10^5$.  At this radius, Comptonization of the stellar
photons switches off suddenly (see below) and bremsstrahlung takes
over as the dominant cooling.

\bigskip
\noindent{\bf 4.6. Dominant Cooling Mechanisms}
\bigskip

The ion temperature in our models is essentially determined by $f$,
and for a large $f\rightarrow 1$, it is close to the virial
temperature.  The electron temperature, which is the temperature that
is relevant for understanding the emission from the system, is however
determined primarily by the cooling processes and the coupling between
ions and electrons.  We investigate in this section which cooling
processes dominate as a function of radius in various accreting
systems.  We remind readers that our flows are assumed to have
equipartition magnetic fields.  These fields allow strong cooling
through thermal synchrotron radiation under the right conditions.  If
such fields are absent ($\beta=1$) or if they somehow do not
participate in cooling, then the results will be quite different.
Note in any case that we do not have nonthermal particles and
therefore there is no nonthermal synchrotron emission in our model.

In Figure 6, we show as a function of $r$ the net cooling rates per
unit volume due to bremsstrahlung ($q_{br}^-+q_{br,C}^-$),
synchrotron ($q_{synch}^-+q_{synch,C}^-$) and soft photon
Comptonization ($q_{*,C}^-$) for various systems.  In each case, we
show the contributions due to these cooling processes along the
critical line $\md_{crit}$ in the $\dot mr$ plane.  The first two
panels are for two accreting black hole models, one corresponding to
$M=10M_\odot$ and the other to $M=10^8M_\odot$.  At large $r\
\sgreat\ 10^4$, we see that the cooling is dominated by
bremsstrahlung.  Recall that this is approximately the region where
the critical accretion rate $\dot m_{crit}$ can be calculated using
the simple relation given in eq (4.1).  At lower radii, the electron
temperature becomes large, $T_e\ \sgreat\ 10^9$K, and the
quasi-relativistic thermal electrons discover new channels of cooling
through synchrotron emission and Comptonization.  Therefore, for $r\
\sles\ 10^2$, synchrotron emission dominates over bremsstrahlung.
This switchover is partially responsible for the deviation of the
critical $\dot m_{crit}$ line from eq (4.1) (see Fig. 1).  Note that
there is very little difference between the $10M_\odot$ case and the
$10^8M_\odot$ case, confirming the scale-free nature of the black hole
solutions.

Figure 7 shows a more detailed representation of the various cooling
processes for an accreting $10M_\odot$ black hole.  For this figure we
scanned the entire $\dot mr$ plane below the critical line, and at
each point we determined which of the four cooling terms, $q_{br}^-$,
$q_{synch}^-$, $q_{br,C}^-$, $q_{synch,C}^-$, dominates.  We have
labeled the zones where these terms dominate as 1, 2, 3, 4
respectively in the figure.  There are two clear trends.  First,
synchrotron-related cooling dominates at lower radii and
bremsstrahlung-related cooling at larger radii as we have already
noted.  Second, we see that the importance of Comptonization increases
at higher values of $\dot m$.  The cooling due to Comptonization
depends on the Compton $y$ parameter which varies sensitively with the
electron scattering optical depth $\tau_{es}$ through the factor $P$
given in Eq (3.20).  The optical depth is essentially proportional to
$\md$ (see eq 2.15) and therefore as $\md$ increases there is a
catastrophic increase in Compton cooling.  Indeed, at all radii
$\sles\ 10^3$, the position of the critical $\md_{crit}$ is determined
almost entirely just by the requirement that the optical depth be of
order unity (which at these electron temperatures is equivalent to the
Compton $y$ parameter being of order unity).

Figure 7 gives a convenient method of gaining a qualitative
understanding of the cooling processes and spectrum of an
advection-dominated accretion flow.  A system with a specific $\dot m$
corresponds to a horizontal track in this figure.  By identifying
which zones are present along the track, and at what radii, it is
possible to obtain a quick idea of the radiation emission from the
system.  Analogous results are presented by Rees et al.  (1982) for
two-temperature plasmas in their ion torus models.  They present their
results as a function of $T_e$ and $r$ for fixed $\dot m$.  In our
model, $T_e$ is not a free parameter but is determined uniquely by the
various processes that are modeled.

Panel c in Fig. 6 shows the results for an accreting neutron star.  We
see that, in this case, the cooling is dominated by the Comptonization
of soft photons over a wide range of radius all the way out to
$r\sim10^5$.  In these models synchrotron cooling is not competitive
even at the smallest radii.  This is because of the lower electron
temperature compared to the black hole models.  At $r\sim10^5$, the
electron temperature is equal to the temperature of the stellar
emission and beyond this radius the soft photon cooling shuts off and
bremsstrahlung cooling takes over.  Actually, at these radii there
will be Compton {\it heating} of electrons which we have not included
in our models.  This could be an important effect (Ostriker et
al. 1976, see also Grindlay 1978 who discusses the effect of
Compton-heating on the thermal pre-heating in X-ray burst sources),
and we hope to include it in future work.

\bigskip
\noindent{\bf 4.7. Stability of the Advection-Dominated Branch}
\bigskip

Two features of our advection-dominated solutions are that they are
optically thin, and that (at large radii at least) they are dominated
by bremsstrahlung cooling.  Accretion flows are normally thermally
unstable precisely under these conditions (Pringle et al. 1973, Piran
1978).  The reason for the instability is that bremsstrahlung cooling
is proportional to the square of the density.  If the flow is
perturbed to a slightly hotter state than the equilibrium, the density
decreases and this leads to less efficient cooling.  The reduced
cooling then causes the disk to become yet hotter and there is a
runaway instability.

The important new feature of our models is that we have included
advection in the equations, whereas the thermal stability analyses
referred to above did not include this effect.  Abramowicz et al.
(1988) (see also Honma et al. 1991) showed that advection causes high
$\dot m$ optically thick flows to become stable even though radiation
pressure dominates in these systems and one would normally expect the
Lightman-Eardley (1974) instability to operate.  More recently, models
of boundary layers calculated by Narayan \& Popham (1993) had
indicated that the thermal instability has a close connection to
advection even in optically thin flows.  In models of CVs, these
authors noticed that the boundary layer becomes optically thin at low
accretion rates, and goes thermally unstable.  As a consequence of the
instability, the gas in the models heats up rapidly to very high
temperatures, but despite this, Narayan \& Popham (1993) were able to
obtain steady state solutions.  A feature of their solutions is that,
once the temperature of the unstable gas approaches the virial limit,
the flow switches to an advection-dominated mode of accretion and a
perfectly reasonable steady state equilibrium flow results.  The
implication is that, even though the accreting gas is thermally
unstable, it is nevertheless able to survive through advection.  We
explore the issue in some detail here.  Because this is an extremely
important and central result of this paper, we discuss the thermal
stability of the various solutions in considerable detail.  An
analogous but briefer discussion can be found in Abramowicz et
al. (1995).

Figure 8a shows some results corresponding to a specific case with
$m=10$, $\md=10^{-3}$, $r=10^3$, $\alpha=0.3$, $\beta=0.5$.  In these
calculations we allowed $f$ to vary over a wide range of values from
$f=10^{-6}$ all the way up to $f\rightarrow1$.  For each value of $f$,
we ignored the energy balance of the ions as described by equation
(3.34), but we solved all the other equations.  In particular, we made
sure that eq (3.35) is satisfied so that the thermal balance of the
electrons is maintained, i.e. the rate at which the electrons are
heated by Coulomb transfer of energy from the ions is exactly
compensated by the rate of cooling of the electrons.  Having done
this, we then checked the thermal balance of the ions by looking at
the terms in eq (3.34).  Recall that the parameter $f$ is defined such
that the advected energy is given by $q^{adv}=fq^+$.  In thermal
balance, the cooling rate of the ions, which is given by $q^{ie}=q^-$,
should be equal to $q^+-q^{adv}=(1-f)q^+$; in other words, the ratio
$q^-/(1-f)q^+$ should be equal to unity.  Any deviation of this ratio
from unity indicates that the ions are not in thermal balance.  The
sense of the deviation, i.e. whether the ratio is greater than or less
than unity, tells us whether the ions will respond by cooling or
heating.

In Fig. 8a we show the calculated values of the ion temperature $T_i$
as a function of the ratio $q^-/(1-f)q^+$.  We obtain a characteristic
S-shaped curve which intersects the equilibrium condition,
$q^-/(1-f)q^+=1$, at three points marked A, B, C.  These three points
represent three equilibrium solutions which satisfy all the equations,
including eq (3.34).  All three solutions are valid solutions of the
equations, but they have different stability properties.

Consider the solution A which has the lowest value of $T_i$.  This is
the standard thin accretion disk solution.  At the position of this
solution, we see that the line describing $T_i$ vs $q^-/(1-f)q^+$
increases upwards to the right.  Let us imagine taking this
equilibrium and perturbing it to a slightly hotter ion temperature
$T_i$, and let us assume that the electron temperature quickly adjusts
so as to maintain eq (3.35).  The system will then move to a new
position on the curve corresponding to the new $T_i$.  Because of the
positive slope of the line, we see that the system will go to a state
with $q^-/(1-f)q^+>1$, i.e. with $q^{ie}+q^{adv}>q^+$.  This means
that the right-hand side of eq (3.34) will be larger than the left
hand side, and the cooling of the ions will dominate over the
heating.  Obviously, the gas will respond by cooling and this means
that $T_i$ will automatically decrease.  Similarly, if we perturb the
system to a lower value of $T_i$ from the equilibrium, then
$q^-/(1-f)q^+$ will decrease below unity and the gas will heat up and
go back towards the equilibrium solution.  Thus, the thin disk
solution A is thermally stable.

Consider next the solution marked B in Fig. 8a.  Here the curve
increases upward to the left.  Therefore, by the same argument as
above, this solution is unstable.  For instance, if we set up an
equilibrium flow corresponding to this solution and perturb it to a
slightly hotter state, then the ion cooling will {\it reduce} relative
to the heating.  The gas will therefore heat up further, and we will
have a classic thermal instability (Pringle et al. 1973, Piran 1978).
The solution B is the standard hot solution (SLE) which has been known
for many years and which has been recognized to be thermally unstable.

Finally, consider the upper most solution marked C with the highest
value of $T_i$.  This is our new advection-dominated branch.  At the
position of this solution, we find that the curve once again rises
upwards to the right, as in solution A.  By the same arguments as
before this solution is therefore stable.  The interesting thing is
that this flow is even more optically thin and more bremsstrahlung
dominated than solution B, and one would normally expect this solution
to be violently unstable.  The secret to its stability is the presence
of the advection term in our equations.

To prove this assertion, we show in panel b a calculation where for
the same system we calculate the ratio $q^-/q^+$.  The point is that
if there is no advection, then we require $q^-=q^+$, and so the ratio
$q^-/q^+$ has to equal unity in equilibrium.  For easy comparison with
previous discussions of the thermal instability in the literature we
include only electron-ion bremsstrahlung cooling in the calculation.
We see that the curve in Fig. 8b satisfies the equilibrium condition
$q^-/q^+=1$ at only two points, identified as A and B.  Solution A is
stable and corresponds to the standard thin accretion disk as before.
Solution B is unstable, again as before.  If such a system were set up
in a state where the ion temperature is hotter than the equilibrium
temperature of solution B, then it would heat up without limit and
would never find a thermally stable state.  This is clearly a
paradoxical situation.  In contrast, if a system is described by Fig.
8a, then, regardless of what initial state it begins with, the flow
can always find a stable solution.  This indicates the great
importance of including advection when one is discussing the thermal
stability of accretion disks.

The above discussion was for a system with a fixed $\md$.  In Fig. 2
we showed how the three branches behave as $\md$ is varied.  There we
plotted $f$ versus $\md$ for each of the branches.  Figure 8c shows
the same branches again, this time showing how the ion temperature
$T_i$ varies along the three solutions.  The three branches are
labeled A, B, C as in the panel a.  Finally, in Fig. 8d we show the
same information as in Figs. 2, 8c in yet another representation where
we plot $\md$ vs the surface density $\Sigma=\rho H$ of the flow.
This kind of representation is commonly used in the literature in
discussions of stability.  We have indicated the two stable branches
by solid lines and the unstable branch by a dotted line.  An analogous
figure can be found in Abramowicz et al. (1995) and Chen (1995).  Note
that at very low $\md\ \sles\ 10^{-6}$, we find $T_i\ \sles\ 10^4$K in
the A branch.  At these low temperatures, the cooling is more
complicated than our simple model because of H recombination, and the
effect of molecules, dust, etc.  We have not tried to include these
opacities in our model because we are at this point interested merely
in demonstrating the basic physics which distinguishes the various
branches.

Finally we would like to emphasize that the systems we are modeling
here do not undergo a thermal limit cycle.  In discussions of the
thermal instability associated with hydrogen ionization (Smak 1984,
Mineshige \& Wheeler 1989) an S curve in the $\md$ vs $\Sigma$
plane is often shown.  Abramowicz et al. (1995) show an analogous S
curve corresponding to their high $\dot m$ optically thick disks
around black holes. In both those cases, the topology of the S is
different from that in Fig. 8d.  As a result, for certain choices of
$\md$, the only solution available in those problems is an unstable
one.  It can be shown that the accretion will then be forced into a
limit cycle behavior where the flow oscillates between the two stable
branches.  In our case, there is no such problem.  Over the ranges of
$\dot m$ and $R$ we consider, there is always at least one stable
solution available and therefore there is no need for a limit cycle.
In fact, for a wide range of $\md$ we actually have {\it two} stable
solutions, which brings up an important question: Which of the two
states does the system choose?  We discuss this issue in the next
section.

We conclude this section with two comments.  First, we note that the
thin accretion disk solution becomes unstable at sufficiently high
$\dot m$ because of the effect of radiation pressure (Lightman \&
Eardley 1974).  Our analysis cannot demonstrate this instability
because we have neglected radiation pressure (for simplicity).  The
analysis of Abramowicz et al. (1988) does include radiation pressure
and they see the effects of this instability as an S curve in the A
branch near the top right corner of Fig. 8d.

Our second comment concerns the viscous stability of the solutions.
In addition to the thermal instability discussed above in detail,
accretion disks are also prone to a secular viscous instability
whenever the slope $d\dot m/d\Sigma$ is negative (e.g. Lightman \&
Eardley 1974, Piran 1978).  Figure 8d shows that all three branches of
solutions we have considered have $d\dot m/d\Sigma>0$ and are
therefore viscously stable, except in very small regions around the
turning points.  The viscous stability of the A and B branches was
already known (e.g. Wandel \& Liang 1991).  We see that our new
advection-dominated branch is also viscously stable.

\vfill\eject
\noindent{\bf 5. When Does Optically Thin Advection-Dominated Accretion
Occur?}
\bigskip

In \S4 we calculated the critical $\md_{crit}$ for optically thin
advection-dominated accretion and presented results for accreting
black holes and neutron stars (Figs. 3--5).  All points on the $\md
r$ plane which lie below the calculated $\md_{crit}(r)$ line have a
stable advection-dominated branch of solutions.  However, over the
entire region between the upper critical line $\md_{crit}$ and the
lower line $\md_{crit}^{'}$ (which is at an extremely low accretion
rate, see eq 4.2), a second solution is also allowed, namely the
standard cooling-dominated thin disk solution.  Which of the two
solutions will an accretion flow actually choose?  In the following
three subsections we suggest three answers to this question, the
first of which is the most conservative and the last the most radical
and sweeping.  The three suggestions are not mutually exclusive.

\bigskip\noindent
{\bf 5.1 Advection-Domination Triggered by the Lightman-Eardley Instability}
\bigskip

Our first argument is based on the well-known fact that the thin disk
solution is unstable at high $\md$ and small $r$ due to the effect of
radiation pressure (Lightman \& Eardley 1974).  Whenever radiation
pressure dominates over gas pressure, a thin accretion disk becomes
both thermally and viscously unstable.  The transition to radiation
pressure domination happens across the following line in the $\md r$
plane (Frank et al.  1992),
$$
\md\approx
4.3\times10^{-3}\alpha^{-1/8}m^{-1/8}r^{21/16},\eqno (5.1)
$$
which is shown in Fig. 9a,b for an accreting $10M_\odot$ and
$10^8M_\odot$ black hole.  The critical advection-dominated line
$\md_{crit}$ is also shown in the figure.  We see that there is a
small triangular-shaped area towards the left of the figure where a
thin disk suffers the Lightman-Eardley instability but where a stable
advection-dominated solution is possible.  We argue that, in this
region of parameter space, accretion has to occur in the
advection-dominated mode since this is the only stable solution
available.  Thus, our most conservative proposal is that the optically
thin advection-dominated solutions discussed in this paper are
restricted to the narrow triangular regions shown in Fig. 9 (plus the
region $\md<\md_{crit}^{'}$, which is at much too low an accretion
rate to be of practical importance).  A proposal similar to this is
widely discussed in the literature, where it is proposed (e.g. SLE)
that a disk which is unstable by the Lightman-Eardley mechanism will
switch to the SLE hot solution.  The main difference in what we
propose is that the relevant hot state is not the SLE solution but
rather our new advection-dominated solution.  Incidentally, note that
our advection-dominated solutions are always gas pressure dominated
and therefore are in no danger from the Lightman-Eardley instability.
This is why it is appropriate for us to neglect radiation pressure in
our models of the advection-dominated state.

As an aside we note that the Lightman-Eardley (1974) instability
operates only if the viscosity coefficient is taken to be proportional
to the total pressure (gas plus radiation pressure).  The instability
is absent if viscosity is proportional only to the gas pressure.
Although such a prescription has been motivated by some earlier work
(e.g. Lightman \& Eardley 1974, Coroniti 1981, Stella \& Rosner 1984),
it is not clear how realistic such a prescription is, especially in view
of the work of Balbus \& Hawley (1991) on magnetic viscosity.

\bigskip\noindent
{\bf 5.2 Effect of Initial Conditions}
\bigskip

As a slightly less conservative proposal, we suggest that an
advection-dominated flow occurs whenever the initial conditions of the
accreting material at the outer radius correspond to an advective
form.  Consider as a specific example the case shown in Fig. 10a which
consists of a $10M_\odot$ black hole accreting at a mass accretion
rate of $\md=10^{-2}$.  The track of the system, shown by a dotted
line in the figure, intersects the critical line at
$r=r_{crit}=10^{5.5}$ (the filled circle).  Suppose the black hole
accretes from a companion star and suppose the outer radius where the
material is injected lies at $r_{out}<r_{crit}$.  If the material
comes in very hot, i.e.  close to virial, or if the material comes in
cold but undergoes an adiabatic shock when it meets the accretion disk
and heats up to the virial temperature, then the initial conditions
will place the gas in the advection-dominated branch.  Since this is a
stable solution, we propose that the flow will continue inwards along
the same branch.  In other words, even though a thin cooling-dominated
disk is allowed the gas does not discover this branch of solutions
because of its initial conditions.  If, however, the accretion begins
with $r_{out}>r_{crit}$, then regardless of whether the initial
conditions are hot or cold, the incoming gas will initially have to
form a thin disk.  This is because $\md>\md_{crit}(r_{out})$ and no
advection-dominated solution is allowed.  Once the flow is trapped in
the thin disk branch at $r=r_{out}$ it will follow this branch all the
way down to a small radius until either it hits the horizon or passes
into the Lightman-Eardley unstable zone.  Thus, in this proposal,
everything depends on initial conditions.

\bigskip\noindent
{\bf 5.3 Advection-Dominated State Preferred Whenever it is Available}
\bigskip

We come now to our most radical suggestion, which we feel is quite
possibly the way nature really works.  We speculate that over much of the
region below the critical line, i.e. whenever $\md<
\md_{crit}(r)$, an accretion flow selects the
advection-dominated branch.  We base this speculation on a feature of
the vertical structure of thin disks which makes cooling-dominated
solutions prone to a surface instability.

A number of researchers have attempted to go beyond the ``one-zone''
solutions that are common in thin accretion disk theory in order to
find detailed solutions for the vertical structure of these disks
(Shaviv \& Wehrse 1986, 1989, Adam et al. 1986).  The most complete
work so far is by Shaviv \& Wehrse (1989) who set up a self-consistent
set of equations for hydrostatic equilibrium, energy generation, and
radiative transfer, and numerically calculated the vertical solution.
One of the surprising results of their work is that for many
reasonable forms of the energy generation law, they find that the
uppermost layers of the disk become thermally unstable so that it is
impossible to find steady state solutions.  We note, however, that the
thermal instability does depend on the details of the viscous energy
generation and the opacity in the optically thin surface layer.  The
explanation of the instability is straightforward.  In any solution
for the vertical structure, the outermost layer (the atmosphere) has
very low density and therefore has poor cooling efficiency.  This
region is also optically thin to the outside.  If viscosity dissipates
any energy in this low density optically thin material, the gas has no
way of radiating the energy and therefore has to heat up.  This is
exactly the thermal instability discussed earlier except that it is
restricted to the outermost layer of the disk at optical depth $\tau\
\sles \ 1$.  Any irradiation of the surface will only enhance the instability.

What happens to the unstable surface gas?  We suggest that it will
keep on expanding until it becomes almost virial.  Once virial, the
gas will be forced to switch to an advection-dominated mode and will
form an advective differentially-rotating accreting corona on top of
the disk.  Moreover, the optical depth of the material will decrease
drastically because its radial velocity will be much greater than the
velocity in the disk.  Once the top layer of the disk has boiled off,
what about the remaining gas in the disk?  A new layer on the surface
of the disk, down to optical depth unity, will next become unstable
and it will take off into the corona.  In this manner the disk surface
will continue to boil away until the entire disk has been converted
into an accreting corona.  The final state is of course what we call
our advection-dominated branch of solutions.  Note that the
instability which we invoke is explicitly a property of the vertical
stratification of the disk and is beyond the scope of the
height-integrated equations we have used in this paper.  This is why
our analysis of \S4.7 showed the thin disk branch to be stable.
Incidentally, since the advection-dominated flow is much more
homogeneous as a function of height (see the solutions in Paper 2) and
since it is anyway insensitive to cooling, it is not expected to be
unstable by this mechanism.

If we accept the above instability, then we believe that any thin disk
which is prone to a vigorous surface instability and which has
$\md<\md_{crit}$ will over a period of time evaporate into a corona.
The critical question is: Under what conditions does the surface
instability act with sufficient speed to ensure complete evaporation
of the thin disk?  This is a difficult question, whose answer depends
on the vertical stratification of the viscous stress and on the
details of the optically thin cooling in the surface layer.  The topic
is beyond the scope of this paper, but we believe there is a real
possibility that, over a wide region below the critical $\md$ lines
which we have calculated in this paper, the true equilibrium flow is
the advection-dominated solution and that no thin disk flows should be
seen.

What does the surface instability do to accretion flows which lie
above the critical line?  In these cases again the surface layers of
the disk will become unstable and will evaporate to form a corona.
However, the entire disk cannot evaporate because there is no
advection-dominated solution which can handle such a large $\md$.  We
suggest therefore that the disk will evaporate only until the corona
has the maximum $\md$ allowed at the given $r$.  The rest of the mass
will remain in the cool disk.  We thus envisage a sandwich structure
where the corona will be maximally advection-dominated, while the
central thin disk will be cooling-dominated.  We obtained solutions
similar to this in Paper 2 where some of our self-similar flows had a
high density equatorial flow coexisting with a very hot low density
advective corona on top.  The role of pairs in such a structure could
be important but is uncertain at this point.

Figure 10b shows schematically how the proposal works for the specific
example considered earlier.  Let us suppose that $\md=10^{-2}$ and
$r_{out}>r_{crit}$.  Let us also assume that evaporation occurs very
rapidly at all radii.  At radii $r>r_{crit}$, the accretion rate lies
above the critical $\md_{crit}(r)$.  Therefore, in this region, we
suggest that we will have a saturated corona with an accretion rate
given by $\md_{cor}=\md_{crit}$, and that the excess material will
flow through the equatorial thin disk with
$\md_{disk}=\md-\md_{crit}$.  As $r$ decreases, the relative fraction
of the mass in the corona increases and that in the disk decreases.
When $r$ reaches the critical value $r=r_{crit}$, all the mass in the
disk will have evaporated into the corona.  Below $r_{crit}$, the
accretion occurs purely in an advection-dominated mode.  Thus, the
flow has three zones: there is a thin disk zone at large $r>r_{crit}$,
a corona above the thin disk over the same range of $r$, and a fully
advection-dominated zone at small $r<r_{crit}$.

Meyer \& Meyer-Hofmeister (1994) proposed a new model of low $\dot m$
CVs which is very similar to the one we describe.  Their systems have
normal thin disks at large radii while at smaller radii they assume
that the flow is entirely in a hot nearly virial state, i.e. in an
advection-dominated state.  By postulating a hole in the accretion
disk at small radii they show that they can explain naturally some
puzzling features in CV observations such as the so-called UV delay.
The identification of a stable advection-dominated solution branch in
this paper provides strong support to the Meyer \& Meyer-Hofmeister
proposal.  At the same time, we view their success in explaining CV
observations as support for our suggestion that the
advection-dominated flow is the most natural and stable form of
accretion possible and that accretion flows will often choose this
form when it is available.

Of course, this is just a qualitative proposal at the present time and
will need to be compared with detailed calculations.  For instance,
the maximum $\md_{crit}$ which the corona can handle will not be
exactly the $\md_{crit}$ which we have calculated in this paper.  This
is because the equatorial thin disk is a source of soft photons which
can cause additional Compton cooling (e.g. SLE).  Also, we need to
show that the time scale on which a thin disk evaporates into the
corona is shorter than the accretion time.  Only then is the surface
instability of any importance.  A quantitative discussion of these
issue is beyond the scope of this paper.

\vfill\eject
\noindent{\bf 6. Summary and Discussion}
\bigskip

Advection-dominated flows occur when the local cooling time scale
becomes longer than the accretion time scale, so that most of the
dissipatively liberated energy is advected inward with the accreting
gas rather than being radiated.  The result is that the gas
temperature becomes almost virial and the luminosity of the system
falls well below the standard $GM\dot M/R_*$ that is normally expected
from accretion.  As we have discussed in Papers 1 and 2,
advection-dominated flows can in principle occur either in systems
with very high mass accretion rates (Begelman 1978), where the photon
diffusion time scale is very long, or in systems with very low mass
accretion rates (Rees et al. 1982), where the local optically thin
cooling time scale becomes very long.  In this paper we concentrate on
the latter case.

Our model of the accretion flow combines the dynamical equations of
Papers 1 and 2 (summarized in \S2) with a comprehensive description of
cooling processes which includes bremsstrahlung, synchrotron, and
Comptonization (\S3).  We model the gas as a two-temperature plasma
(following SLE) where the ions and electrons are allowed to have
different temperatures, determined by individual thermal balance
equations for each species.  We consider accretion onto black holes
and neutron stars, the distinction being that in one case all the
advected energy disappears through the horizon whereas in the other
the energy is thermalized and reradiated at the stellar surface.  In
our models, we assume that the flows contain roughly equipartition
strength magnetic fields whose strength is estimated through a
parameter $\beta$ (see eqs 2.6, 2.10).  Hot electrons can cool through
synchrotron emission in the magnetic field.  The field may also be the
agency whereby angular momentum is transported, but we do not model
this process in detail.  Instead we use a standard $\alpha$
prescription for viscosity, where the parameter $\alpha$ is assumed to
include the effects of hydrodynamic turbulence/convection and magnetic
stresses.  Occasionally, in the literature, radiative viscosity is
discussed in the context of accreting black holes and neutron stars
(e.g. Loeb \& Laor 1992, Miller \& Lamb 1993).  As we show in Appendix
B, this viscosity is not important for the flows considered in this
paper.

The main results of our calculations are described in \S4.  By
numerically solving the equations we find that the model gives three
distinct branches of equilibria (\S\S4.2, 4.7), two of which are
stable and one of which is thermally unstable.  One of the stable
branches is the standard cooling-dominated thin accretion disk
solution which is much discussed in the literature (Shakura \& Sunyaev
1973, Novikov \& Thorne 1973, Lynden-Bell \& Pringle 1974).  The
unstable branch corresponds to a hot optically thin solution which is
again much discussed in the literature (SLE, Luo \& Liang 1994, and
references therein).  The third branch, which corresponds to a hot,
optically thin, advection-dominated flow, is a new solution which
comes out of our analysis and which is the principal focus of this
paper.  This solution is thermally stable.  Moreover, as we
demonstrate in \S4.7, it is the introduction of advection which allows
the solution to be stable despite being optically thin.  This branch
has also been discussed by Abramowicz et al. (1995) and Chen (1995)
who consider a simplified version of the theory presented here (see
\S4.1).

Is our new solution branch really new?  Certainly, to the extent that
we have for the first time included advection and treated the dynamics
of the flow consistently, the advection-dominated solution described
in this paper and in Abramowicz et al. (1995) is new.  But, even more
fundamentally, it is our impression that the existence of {\it two}
hot solutions (only one of which is thermally stable) was not
appreciated until now.  Hot solutions have been quite popular for many
years, beginning with the important work of SLE, but these solutions
are calculated under a local cooling assumption, $q^-=q^+$.  As a
consequence, the calculations invariably produce the thermally
unstable solution branch mentioned above.  The ion torus model of Rees
et al. (1982) comes closer to our new advection-dominated solution
since the model specifically avoids the condition $q^-=q^+$.  However,
their paper does not carry out a complete analysis of the dynamics and
it is not clear if the proposed solution is in fact our
advection-dominated solution.  Also, Rees et al. do not explain the
relationship between their solution and the previously known hot SLE
solutions.  Indeed, in much of the accretion literature, the ion torus
is considered to be a variant of the SLE hot solution (e.g. Frank et
al. 1992, p233) rather than as an independent solution branch.

Using our model, we have explored in some detail the properties of the
optically thin advection-dominated branch of solutions.  For each
choice of the stellar mass $M$, the viscosity parameter $\alpha$, and
the equipartition parameter $\beta$ (eq 2.6), we have determined a
critical line $\md_{crit}(r)$ in the $\md r$ plane.  Here,
$\md\equiv\dot M/\dot M_{Edd}$ is the accretion rate scaled to the
Eddington rate with an efficiency $\eta_{eff}=0.1$ (see eq 2.13) and
$r=R/R_{Schw}$ is the radius scaled to the Schwarzschild radius (eq
2.14).  The critical line $\dot m_{crit}(r)$ gives at each $r$ the
maximum accretion rate above which the optically thin
advection-dominated branch of solutions terminates.  Figures 3-5 show
our results for accreting black holes of mass $10M_\odot$ and
$10^8M_\odot$ and for neutron stars.  The critical lines are to be
interpreted as follows.  Any given system with a specified accretion
rate $\md$ corresponds to a horizontal line on the diagram.  If the
system has a high accretion rate so that $\md$ lies entirely above the
critical line then it cannot be advection-dominated.  However, if part
of the flow lies below the critical line, then the flow is allowed to
be advection-dominated over that range of $r$.

In addition to the critical line $\md_{crit}$ discussed above and
shown in Figures 3--5, we show that there is a second limit, which we
designate $\md_{crit}^{'}$ and which is located at an extremely low
accretion rate, $\md_{crit}^{'}\ \sles\ 10^{-10}$ (see eq 4.2).  The
two critical lines divide the $\md r$ plane into three distinct zones.
Above $\md_{crit}$, there is no advection-dominated solution.  Below
$\md_{crit}^{'}$, the thin disk solution is unstable, and the
advection-dominated flow is the only solution allowed.  Between the
two critical lines, however, and this covers a very wide range of
astrophysically interesting parameters, both the advection-dominated
solution and the standard cooling-dominated thin disk solution are
apparently allowed.  Which solution does nature actually pick?

We discuss this issue in \S5 and argue that whenever the thin disk
solution is unstable for any reason the gas will automatically go into
the advection-dominated mode.  In certain regions of the $\md r$ plane
(see eq 5.1) a cooling-dominated thin disk becomes radiation pressure
dominated and in consequence is unstable if the viscosity is
proportional to the total pressure (Lightman \& Eardley 1974).  In
these regions we believe that the flow will automatically switch to
the advection-dominated solution.  A similar idea has been widely
discussed in the literature (e.g. SLE), except that in most previous
discussions a thin disk which is unstable to the Lightman-Eardley
instability is assumed to switch to an SLE-type hot solution, which is
itself unstable.  The new feature here is that we suggest that the
final state is our advection-dominated hot state.

In addition to the well-known Lightman-Eardley instability, we argue
that, even in other regions of the $\md r$ plane, a thin disk has a
different kind of thermal instability which operates on its surface
layers and may cause the disk to evaporate in the vertical direction.
We therefore speculate that thin disks are probably unstable over much
of the region below the critical line $\md_{crit}(r)$ and that many
flows below this line are advection-dominated.  Further, we argue that
even systems with $\md>\md_{crit}$ may have some part of the mass
accretion occurring in an advection-dominated corona.  These ideas are
summarized in Fig. 10, and are similar to suggestions made recently by
Meyer \& Meyer-Hofmeister (1994) in the context of accretion disks in
CVs.

Thermally unstable accretion disks with multiple equilibria have been
discussed in other contexts (Smak 1984, Abramowicz et al. 1988,
Mineshige \& Wheeler 1989), and in most cases it is argued that they
undergo a thermal limit cycle.  The limit cycle is used to explain the
observed variability of the systems.  Although we have multiple
equilibria, the flows described in this paper do not have any tendency
to set up a limit cycle, as we discuss in \S4.7.  Note, however, that
at the higher accretion rates considered by Abramowicz et al. (1988),
limit cycles are indeed a possibility, but their discussion refers to
different solution branches than those considered here.

Generally we find that advection-dominated flows occur for mass
accretion rates ${\dot m} \ \sles\ 10^{-1}-10^{-3}$ for the systems we
have considered.  The results however depend on the value of $\alpha$
and to a lesser extent $\beta$.  There is also a large difference
depending on whether the accreting star is a black hole or a neutron
star (see Figs. 3--5), as anticipated in some previous papers
(e.g. Sunyaev \& Titarchuk 1980, Sunyaev et al. 1991ab, Liang 1993).  At
small $r$, we obtain a critical accretion rate of $\dot
m_{crit}\sim\alpha^2$ for black hole systems, and $\dot
m_{crit}\sim0.1\alpha^2$ for neutron star systems.

All of our advection-dominated solutions have certain common features.
The ion temperature is always close to virial at all radii.  Since the
flow is advection-dominated, most of the energy which is released by
viscosity is retained in the gas, and the ions therefore have to heat
up to close to the virial temperature.  Our solutions are thus similar
to the ion torus model of Rees et al. (1982), except that our flows
are more nearly spherical than toroidal (see Paper 2).  At large
radii, the electrons are in near thermal equilibrium with the ions and
are therefore virial.  However, at radii $r\ \sles\ 10^3$ when the ion
temperature exceeds $10^9$K, the electron temperature deviates from
the ion temperature.  This is because, above $10^9$K, the ion-electron
coupling via Coulomb collisions becomes weak and at the same time the
electrons are able to cool by a variety of processes including
synchrotron emission and Compton cooling.  The efficient cooling keeps
the electrons at a roughly constant temperature independent of $r$,
whereas the ions decouple from the electrons and continue to become
hotter as $r$ decreases.

The two-temperature feature of our solutions is largely a consequence
of our assumption that the only coupling between ions and electrons is
via Coulomb scattering.  If there are non-thermal modes of energy
transfer from ions to electrons (Phinney 1981), for example by the
collective mechanism discussed by Begelman \& Chiueh (1988, hereafter
BC), then the results could be strongly modified.  We discuss this
important issue in Appendix A.  The BC mechanism makes use of wave
interactions in regions of large density perturbations to transfer
energy directly from waves to electrons.  We consider two cases in
Appendix A.  If the density perturbations are caused by a normal
Kolmogorov-like turbulent cascade, then we show that for most cases of
interest, the heating of electrons via Coulomb collisions is stronger
than that due to the BC mechanism. When either $\alpha$ or mass accretion
rate is very small, heating by the BC mechanism may dominate over
the Coulomb heating. Even in this case, however, the viscous heating
rate of ions is much larger than the BC heating rate of electrons
except for very small values of $\alpha$, which suggests that the
two-temperature assumption is justified nearly in all cases. Therefore,
it is appropriate to neglect non-thermal heating of electrons.
In the alternate case, where most of the dissipation occurs in
collisionless shocks, we still find that the BC mechanism is unimportant
so long as the accretion star is a stellar-mass black hole.
However, if the star is
a supermassive black hole, then it appears that the BC non-thermal
heating of electrons wins over Coulomb heating.  Therefore, depending
on the nature of the dissipation, our results may or may not be valid
for AGN.

One of our interesting results is that black hole models are almost
scale-independent.  For the whole range of masses from $M=10M_\odot$
to $10^8M_\odot$ we find that the position of the critical line
$\md_{crit}(r)$ as well as the dependences of the ion and electron
temperatures on scaled radius $r$ are nearly the same.  Individual
physical quantities (volume density, magnetic field strength, etc.)
do vary by large factors from stellar mass to supermassive black holes
but the electron temperatures remain virtually identical.  This
scale-free behavior is true only for the advection-dominated branch of
solutions.  Cooling-dominated thin disks are known to depend on the
mass of the central star, e.g. the surface temperature varies as
$M^{-1/4}$.  The scale-free nature of advective flows may explain some
of the similarities that are observed between galactic black hole
X-ray binaries and active galactic nuclei.  The bremsstrahlung
emission at least in the two cases must have similar characteristics.
The synchrotron radiation however will be at very different
frequencies, since the self-absorption frequency $\nu_c$ (equation
3.15) scales as $M^{-1/2}$.

Our calculations indicate that there are strong differences between
advection-dominated accretion flows around black holes and those
around neutron stars (compare Figs. 3 and 5).  The difference arises
primarily because in the latter case all the advected energy is
re-radiated as blackbody radiation from the stellar surface, whereas
in the former the energy is lost into the hole.  This point has been
emphasized before (e.g. Sunyaev et al. 1991ab, Liang 1993) and leads to
several effects.

\noindent
(i) The spectrum of an accreting neutron star will have an additional
cool component due to the radiation from the star, which will be
missing in an accreting black hole.

\noindent
(ii) Cooling is more effective in the neutron star case because the
soft photons from the star Compton-cool the electrons in the accreting
gas very effectively.  We see clear evidence for this in our
calculations, thereby confirming previous suggestions of the effect
(Sunyaev et al. 1991ab, Liang 1993).

\noindent
(iii) As a result of (ii), the critical line for an accreting neutron
star lies at a much lower value of $\md$ than for a black hole.
Because of this shift, we suggest that a significantly larger fraction
of black hole binaries in the Galaxy may be in an advection-dominated
mode compared to neutron star binaries.

\noindent
(iv) One of the features of the advection-dominated branch is that the
radiation luminosity is low (Phinney 1981, Rees et al. 1982).
Combining this with the previous points, we suggest that the majority
of black hole binaries may be unusually dim for their accretion rate.
This may have some implications for understanding transient X-ray
sources, many of which are believed to be accreting black holes.  It
may also explain why black hole X-ray binaries typically have similar
luminosities as neutron star binaries even though the central stars in
the former are believed to be much more massive.  Figure 11
illustrates quantitatively the difference between accreting black
holes and neutron stars.  The two panels show the luminosities of the
two systems as functions of $\dot m$.  We see that the luminosity of
the black hole accretor drops rapidly with decreasing $\dot M$, giving
extremely low radiative efficiency for $\dot m
\ll\alpha^2$.  Roughly, the luminosity scales as $L\sim 0.1\dot M c^2
(\dot m/\alpha^2)$.  In contrast, an accreting neutron star has a more
or less constant efficiency of $\sim20\%$, independent of $\dot m$,
i.e. $L\sim 0.2\dot M c^2$.  Of course, at low $\dot m$, the accretion
flow is highly advection-dominated in the neutron star case as well,
as shown by the dotted line in Fig. 11.  However, all the advected
energy is re-radiated from the surface of the star, and therefore the
net efficiency is high.

\noindent
(v) The electron temperatures in the two cases are different.  For
$\dot m\sim\dot m_{crit}$, accreting black holes reach
$T_e\sim10^9-10^{9.5}$K at small radii, whereas accreting neutron star
flows saturate at $T_e\sim10^{8.5}-10^9$K.  For $\dot m\ll\dot
m_{crit}$, the temperatures in the black hole systems are even higher,
$T_e\ \sgreat\ 10^{10}$ K, but the neutron star systems hardly ever
exceed $T_e\sim10^9$K.  As before, the difference between the two
kinds of systems is because of the additional cooling in the neutron
star case due to Comptonization of the stellar photons.  A robust
prediction therefore is that black hole systems will have harder
spectra, as appears to be the case in some of the observations.

\noindent
(vi) Given the high electron temperatures in our models, we may expect
pair effects to become important.  It turns out that this is not an
issue for any of the models described in this paper.  Our estimates of
the equilibrium positron fraction, calculated using the relations given
in White \& Lightman (1989, see also Kusunose \& Takahara 1989), are
never larger than $\sim 10^{-3}$.  However, at higher accretion rates
than those we have considered, i.e. for $\md>\md_{crit}$, the optical
depth will increase rapidly and it is quite possible that various
kinds of pair instabilities may occur.  Based on the fact that our
black hole models are invariably hotter than the neutron star models,
we think that perhaps pair effects will be most significant for
accreting black holes at high $\md$.  We further feel that accreting
neutron stars may never reach the required temperatures for copious
pair production at any value of $\md$.  Observationally, we note that
pair annihilation lines have been seen only from black hole candidates
and not from any confirmed neutron star system.

In discussions of AGN, a spherical accretion model has been discussed
sometimes in the literature (e.g. McCray 1979, Ipser \& Price 1983,
Melia 1994).  In this model, the angular momentum of the accreting gas
is completely neglected, a rather severe approximation.  Although our
advection-dominated solution is quasi-spherical (see Paper 2) it does
include differential rotation, viscosity, angular momentum transport,
etc., all in a self-consistent fashion.  We may therefore describe our
models as more realistic and self-consistent versions of the idealized
spherical accretion models discussed in the past.

In future papers we intend to discuss the radiation emitted by our
advection-dominated flows and to compare model spectra with
observations of low-$\dot m$ black hole and neutron star systems.  The
information in Fig. 7 indicates what we may qualitatively expect for
the spectrum.  Since the bulk of the emission in an accretion flow
comes from small radii, the spectrum will be dominated by synchrotron
emission and Comptonization of the synchrotron photons.  The primary
synchrotron emission in the models will give a peak in the spectrum at
the self-absorption frequency $\nu_c$ (equation 3.15).  The position
of the peak depends on the particular values of the various
parameters, $M$, $\dot M$, $\alpha$, $\beta$, but roughly we expect
the synchrotron peak to occur at $\nu\sim 10^{15}$ Hz for a
stellar-mass object and at $\nu\sim10^{11}$ Hz for a $10^8M_\odot$
black hole.  The Comptonization of the synchrotron emission will
produce one or more bumps in the spectrum at higher frequencies, or a
near power-law spectrum if the Compton $y$-parameter is large enough.
Rather interestingly, we see from Fig. 7 that Comptonization is quite
important for a range of accretion rates from $\dot m\sim\dot
m_{crit}$ down to $\dot m\sim 0.01\dot m_{crit}$.  Finally, the
spectrum will have a peak at a frequency $\nu\sim 10^{20}$ Hz (i.e. at
a photon energy of order a few hundred keV), due to bremsstrahlung
emission by the hot electrons.

\bigskip\noindent
{\it Acknowledgements}: We thank D. Barret, J. Grindlay, J.-P. Lasota,
E. Liang, R. Mahadevan and J. Ostriker for useful discussions, and
M. Abramowicz, R. Blandford, D. Eardley, A. Lightman, S. Phinney,
M. Rees and S. Shapiro for comments.  We are grateful to our second
referee, Dr. R. E. Taam, for prompt arbitration after the paper was
rejected by the first referee, and for making a number of very constructive
suggestions.  This work was supported in part by grant AST9148279 from
the National Science Foundation.

\vfill\eject
\noindent{\bf Appendix A: Non-Thermal Coupling of Ions and Electrons}
\bigskip

Throughout this paper we have assumed that the only coupling between
ions and electrons is via Coulomb collisions.  This assumption leads
to a two-temperature plasma in the inner regions of our solutions, and
it is the two-temperature feature which allows our flows to be
advection-dominated.  However, it has been argued (cf. Phinney 1981)
that, in addition to Coulomb collisions, many other non-thermal
processes may couple ions and electrons efficiently and these
processes could potentially eliminate the two-temperature structure.
This is an important issue for our work since most of the properties
of our solutions are a consequence of the two-temperature nature of
the plasma.

In the astrophysical literature, the only work we are aware of
discussing a specific non-thermal mechanism to heat electrons is that
of Begelman \& Chiueh (1989, hereafter BC).  BC start with the
interesting fact that a two-temperature plasma has the following
unusual ordering of lengthscales,
$$
\lambda_{ce}<\lambda_{De}\ll\lambda_{Di}<\lambda_{ci},\eqno (A1)
$$
where $\lambda_{De}, ~\lambda_{Di}$ are the electron and ion Debye
lengths, and $\lambda_{ce}, ~\lambda_{ci}$ are the respective
gyroradii.  BC then show that, under these conditions, plasma waves
traveling through a suitably turbulent zone can pump energy directly
into the electrons through collective motions.  The energy transfer
via this mechanism may, in principle, occur rapidly enough to
eliminate the two-temperature structure of the plasma.

We calculate here the rate of heating of electrons via the BC
mechanism for the specific flows considered in this paper, and compare
this rate to the rate of energy tranfer from ions to electrons via
Coulomb collisions.  For simplicity, we set $\beta=0.5$ (equipartition
fields), $f=1$ (fully advection-dominated flow), and correspondingly
choose $c_1=0.489, ~c_2=0.429, ~c_3=0.326, \ep=\epp=0.565$.  Also, we
set $T_e\ll T_i$ in equation (2.16), so that $T_i$ is equal to the
right-hand side of this equation.

 From the expression for $q^+$ in equation (2.15), we find that the
rate at which an ion is heated by viscous dissipation is given by
$$
\left({dE_i\over dt}\right)_{vis}=8.95\,\alpha m^{-1}r^{-5/2}
{}~{\rm erg\,s^{-1}}.\eqno (A2)
$$
Assuming that the ions are non-relativistic and the electrons are
relativistic, equation (3.3) for the rate of heating of electrons by
Coulomb collisions simplifies to
$$
\left({dE_e\over dt}\right)_{Coul}=5.9\,\alpha m^{-1}\mda
r^{-5/2}T_{10}^{-1} ~{\rm erg\,s^{-1}},\eqno (A3)
$$
where we have written $T_e=10^{10}T_{10}\,$K.  Recall that, because of
the efficient cooling of the electrons, any energy transferred to the
electrons is immediately radiated.  Therefore, for an
advection-dominated solution, we require
$(dE_e/dt)_{Coul}<(dE_i/dt)_{vis}$.  From equations (A2) and (A3),
this gives the condition
$$
\mda<\mda_{crit}\sim1.5T_{10}.\eqno (A4)
$$
This scaling is consistent with the detailed numerical results for
$\dot m_{crit}$ shown in Fig. 3 for the inner regions of accreting
black holes.

According to BC, the rate of heating of electrons when their
instability is fully developed is given by
$$
\left({dE_e\over dt}\right)_{BC}\sim f_{BC}{m_iv_{ti}^3\over
\lambda_{Di}}\left({v_A\over c}\right)^2\left({\lambda_{ci}\over
L}\right)^{9/2},\eqno (A5)
$$
where $f_{BC}$ is the filling factor of the plasma where the kind of
strong turbulence needed for the instability is present, $m_i$ is the
mass of the ions (which we take to be protons), $v_{ti}$ is the ion
thermal velocity, $v_A$ is the Alfven velocity, and $L$ is the local
density gradient scale.  For the self-similar scalings given in
equation (2.15), we have
$$
v_{ti}=8.56\times10^9\,r^{-1/2} ~{\rm cm\,s^{-1}},\eqno (A6)
$$
$$
\lambda_{Di}=8.84\times10^{-4}\,\alpha^{-1/2}m^{1/2}\mda^{-1/2}
r^{1/4} ~{\rm cm}, \eqno (A7)
$$
$$
v_A=1.21\times10^{10}\,r^{-1/2} ~{\rm cm\,s^{-1}},\eqno (A8)
$$
$$
\lambda_{ci}=3.13\times10^{-3}\,\alpha^{-1/2}m^{1/2}\mda^{-1/2}
r^{3/4} ~{\rm cm}. \eqno (A9)
$$
The values of $f_{BC}$ and $L$ depend on the particular turbulence
scenario we consider.  We discuss two limits below.

\bigskip
\noindent{\bf Kolmogorov-Like Turbulent Cascade}
\medskip

Let us suppose we have a space-filling turbulent cascade, and let us
assume that it satisfies the Kolmogorov scaling, with an outer scale
equal to the disk thickness $H$.  Then, following equation (5.9) of
BC, the gradient scale $L$ satisfies
$$
\left({L\over\lambda_{ci}}\right)\sim100\,\alpha^{1/6}m^{-1/6}
\mda^{-1/6}r^{-1/12}.\eqno (A10)
$$
The filling factor $f_{BC}$ is potentially equal to unity in the case
of a turbulent cascade, but we retain it as a factor since it is
possible that only a fraction of the volume actually satisfies all the
conditions required for the BC instability.  (For instance, subsonic
turbulence is usually incompressible and it is not clear that it will
have sufficient density contrasts for the BC mechanism to operate.)
Then, the rate of heating of electrons by the BC mechanism is given by
$$
\left({dE_e\over dt}\right)_{BC}\sim 0.21\,\alpha^{-1/4}m^{-5/4}
\mda^{-1/4}r^{-25/8} ~{\rm erg\,s^{-1}}.\eqno (A11)
$$
Now, BC have shown that their instability operates only if the scale
$L$ satisfies a particular inequality (cf. equation 5.6 in their
paper).  For our choice of $\beta$, the relevant condition is
$$
1<{L\over \lambda_{ci}}<2.2r^{1/2}.\eqno (A12)
$$
Comparing this condition with equation (A10), we see that the BC
instability operates only for radii $r$ greater than a certain limit:
$$
r>r_{crit}\sim 690\,\alpha^{2/7}m^{-2/7}\mda^{-2/7}.\eqno(A13)
$$
Let us substitute $r=r_{crit}$ into equation (A11).  We then find that
the electron heating rate at the optimum radius satisfies
$$
{(dE_e/dt)_{BC,max}\over(dE_e/dt)_{Coul}}\sim6.0\times10^{-4}\,f_{BC}
\alpha^{-10/7}m^{-1/14}\mda^{-15/14}.\eqno (A14)
$$
The right-hand side of (A14) is less than unity under most conditions
of interest for our solutions.  This shows that Coulomb heating
usually dominates.  Only for very small values of $\alpha$ or very
small mass accretion rates does the BC heating become competitive with
Coulomb heating. Even when the heating ratio (A14) becomes comparable
to or larger than unity, the ratio $(dE_i/dt)_{vis}/(dE_e/dt)_{BC}$ is
usually much larger than unity except when
$\alpha<4\times 10^{-3}$. In other words, unless $\alpha$ becomes
very small, the viscous heating of ions dominates over the
Begelman-Chiueh heating and a two-temperature plasma is very likely.
We note that this conclusion is only very weakly dependent on $m$ and
${\dot m}$.

\bigskip
\noindent{\bf Collisionless Shocks}
\medskip

The other possibility is that the turbulent gas may have some fraction
of the dissipation occurring through collisionless shocks.  In the
case of shocks, we expect the gradient scale $L$ to be of order the
ion gryroradius $\lambda_{ci}$, so that the condition (A10) is easily
satisfied.  To calculate the filling factor of the shock fronts, we
note that the volume rate of dissipation of energy within a shock is
approximately
$$
q_{sh}\sim{\rho v_A^2\over \lambda_{ci}/v_A}\sim2.6\times10^{28}
\alpha^{-3/2}m^{-3/2}\mda^{3/2}r^{-15/4} ~{\rm erg\,cm^{-3}\,s^{-1}}.
\eqno (A15)
$$
Since the total dissipation rate per unit volume is $q^+$, we
can calculate the maximum filling factor $f_{BC,max}$ of the shocks by
equating $f_{BC,max}q_{sh}$ to $q^+$:
$$
f_{BC,max}\sim2.3\times10^{-8}\,\alpha^{5/2}m^{-1/2}\mda^{1/2}r^{-1/4}.
\eqno (A16)
$$
Let us write $f_{BC}=\delta f_{BC,max}$, so that the actual volume
fraction which participates in the BC mechanism is only a fraction
$\delta$ of the maximum allowed.  Then we obtain
$$
{(dE_e/dt)_{BC}\over(dE_e/dt)_{Coul}}\sim0.74\,\delta
\alpha^2m r^{1/2} T_{10}.\eqno (A17)
$$
We see that the BC mechanism may be quite important in the case of shocks,
whereas it is irrelevant in the case of the Kolmogorov cascade
considered earlier.  Note the curious fact that $\dot m$ does not
enter in equation (A17).  For $\alpha\sim 0.1$ and $r\
\sles\ 10^2$ (cf. the two-temperature zone in Fig. 3), we see that the
BC mechanism via shocks becomes more important than Coulomb transfer
if $\delta m>14$.  We have no idea what a reasonable estimate of
$\delta$ may be, but we see that even if $\delta\sim1$, a
two-temperature gas is allowed for flows around stellar-mass black
holes.  Thus, the models developed in this paper are likely to be
valid for Galactic X-ray binaries.  However, it appears that these
solutions may run into difficulties for AGN models, unless $\delta$ is
extremely small.  A small $\delta$ requires either negligible
dissipation through shocks or an inefficient BC mechanism in the
vicinity of shocks.

\vfill\eject
\noindent{\bf Appendix B: Radiative Viscosity}
\bigskip

In our models we have assumed an $\alpha$ form of viscosity which
implicity assumes that the viscous stress has a radius-independent
self-similar form.  Such a viscosity could arise from magnetic
stresses or turbulent stresses (Shakura \& Sunyaev 1973).  There has
been some discussion in the literature of radiative viscosity due to
the scattering of photons off electrons, both in the context of black
hole flows (Loeb \& Laor 1992) and neutron star flows (Miller \& Lamb
1993).  One feature of radiative viscosity is that the effective
$\alpha$ due to it depends on the radius.  Therefore, if this
viscosity is important, we will need to modify the scalings we have
given in equation (2.15).  We discuss here the magnitude of
$\alpha_{rad}$ due to radiation drag for the black hole and neutron
star problems.

Because we have flows which are optically thin to infinity, especially
at low $\dot m$, some of the standard formulae in the literature for
the radiative viscosity are not appropriate.  If the luminosity
of the accreting system is $L$, then we expect the radiation density
$u_{ph}$ at radius $R$ to be of order
$$
u_{ph}\sim {L\over4\pi R^2c}={\eta\dot M c\over4\pi R^2},\eqno (B1)
$$
where $\eta$ is the radiative efficiency of the accretion flow.  The
probability per unit time that a given photon scatters is
$n_e\sigma_Tc$ and on average the post-scattered photon will have a
tangential velocity $\sim\Omega R$.  Thus, the rate at which angular
momentum is removed from the accreting gas per unit volume per unit
time is
$$
\dot J\sim R{u_{ph}\Omega R\over c^2}n_e\sigma_Tc
\sim{L\Omega n_e\sigma_T\over4\pi c^2}.\eqno (B2)
$$
The effective $\alpha_{rad}$ due to this radiative viscosity is
obtained by dividing $\dot J$ by the gas pressure $p$.  Thus
$$
\alpha_{rad}\sim {\eta\dot M\Omega n_e\sigma_T\over 4\pi p}.\eqno (B3)
$$

As in Appendix A, let us set $\beta=0.5$, $f\rightarrow1$ and use the
corresponding values of $c_1-c_3$.
As described in the main paper, the efficiency of our
advection-dominated flows is approximately given by $\eta\sim0.1(\dot
m/\alpha^2)$ in the case of a black hole and $\eta\sim0.2$ in the case
of a neutron star.  We thus find
$$
\alpha_{rad}\sim0.8\alpha^2\mda^2r^{-1/2}\eqno (B5)
$$
in the case of an accreting black hole, and
$$
\alpha_{rad}\sim1.6\alpha^2\mda r^{-1/2}\eqno (B6)
$$
in the case of an accreting neutron star.  We expect $\alpha^2\ \sles\
0.1$.  Also, for advection-dominated flows, $\dot m/\alpha^2<1$ for a
black hole and $\dot m/\alpha^2<0.1$ for a neutron star.  Therefore,
radiative viscosity is not likely to be important for our
optically-thin advection-dominated flows except perhaps very close to
the center.

\vfill\eject
\noindent{\bf References}
\bigskip

\ref{Abramowicz, M., Czerny, B., Lasota, J. P., \& Szuszkiewicz, E. 1988, ApJ,
332, 646}

\ref{Abramowicz, M., Chen, X., Kato, S., Lasota, J. P., \& Regev, O. 1995,
ApJ, Jan 1}

\ref{Adam, S., Stoerzer, H., Wehrse, R., \& Shaviv, G. 1986, A\&A, 193, L1}

\ref{Balbus, S. A. \& Hawley, J. F. 1991, ApJ, 376, 214}

\ref{Begelman, M. C. 1978, MNRAS, 184, 53}

\ref{Begelman, M. C. \& Chiueh, T. 1988, ApJ, 332, 872}

\ref{Begelman, M. C. \& Meier, D. L. 1982, ApJ, 253, 873}

\ref{Chen, X. 1995, preprint}

\ref{Chen, X. \& Taam, R. E. 1993, ApJ, 412, 254}

\ref{Clayton, D. D. 1983, Principles of Stellar Evolution and Nucleosynthesis,
(Chicago: The University of Chicago Press)}

\ref{Coroniti, F. V. 1981, ApJ, 244, 587}

\ref{Dermer, C. D., Liang, E. P., \& Canfield, E. 1991, ApJ, 369, 410}

\ref{Eardley, D. M., Lightman, A. P., Payne, D. G., \& Shapiro, S. L.
1978, ApJ, 224, 53}

\ref{Eggum, G. E., Coroniti, F. V., \& Katz, J. I. 1988, ApJ, 330, 142}

\ref{Frank, J., King, A., \& Raine, D. 1992, Accretion Power in Astrophysics
(Cambridge, UK: Cambridge University Press)}

\ref{Grindlay, J. E. 1978, ApJ, 221, 234}

\ref{Honma, F., Matsumoto, R., \& Kato, S. 1991, PASJ, 43, 147}

\ref{Hubeny, I. 1990, ApJ, 351, 632}

\ref{Ipser, J. R. \& Price, R. H. 1983, ApJ, 267, 371}

\ref{Kusunose, M. \& Takahara, F. 1989, PASJ, 41, 263}

\ref{Liang, E. P. 1988, ApJ, 334, 339}

\ref{Liang, E. P. 1993, Proc. Compton Gamma-Ray Observatory Workshop}

\ref{Liang, E. P. T. \& Thompson, K. A. 1980, ApJ, 240, 271}

\ref{Lightman, A. P. \& Eardley, D. M. 1974, ApJ, 187, L1}

\ref{Loeb, A. \& Laor, A. 1992, ApJ, 384, 115}

\ref{Luo, C. \& Liang, E. P. 1994, MNRAS, 266, 386}

\ref{Lynden-Bell, D. \& Pringle, J. E. 1974, MNRAS, 168, 603}

\ref{Mahadevan, R., Narayan, R., \& Yi, I. 1994, in preparation}

\ref{McCray, R. 1979, in {\it Active Galactic Nuclei}, ed. C. Hazard \&
S. Mitton (Cambridge, UK: Cambridge University Press)}

\ref{Melia, F. 1994, ApJ, 426, 577}

\ref{Melia, F. \& Misra, R. 1993, ApJ, 411, 797}

\ref{Meyer, F. \& Meyer-Hofmeister, E. 1994, A\&A, 288, 175}

\ref{Miller, M. C. \& Lamb, F. K. 1993, ApJ, 413, L43}

\ref{Mineshige, S. \& Wheeler, J. C. 1989, ApJ, 343, 241}

\ref{Muchotrzeb, B. \& Paczy\'nski, B. 1982, Acta Astron., 32, 1}

\ref{Narayan, R. \& Popham, R. 1993, Nature, 362, 820}

\ref{Narayan, R. \& Yi, I., 1994, ApJ, 428, L13 (Paper 1)}

\ref{Narayan, R. \& Yi, I., 1995, ApJ, in press (Paper 2)}

\ref{Novikov, I. D. \& Thorne, K. S. 1973, in Blackholes ed. C. DeWitt
\& B. DeWitt (New York: Gordon and Breach)}

\ref{Ostriker, J. P., McCray, R., Weaver, R., \& Yahil, A. 1976, ApJ, 208,
L61}

\ref{Pacholczyk, A. G. 1970, Radio Astrophysics (San Francisco: Freeman)}

\ref{Paczy\'nski, B. \& Bisnovatyi-Kogan, G. 1981, Acta Astr., 31, 283}

\ref{Paczy\'nski, B. \& Wiita, P. J. 1980, A\&A, 88, 23}

\ref{Phinney, E. S. 1981, in Plasma Astrophysics. ESA SP-161,
Paris: European Space Agency}

\ref{Piran, T. 1978, ApJ, 221, 652}

\ref{Popham, R. \& Narayan, R. 1995, ApJ, in press}

\ref{Popham, R., Narayan, R., Hartmann, L., \& Kenyon, S. 1993, ApJ,
415, L127}

\ref{Pringle, J. E., Rees, M. J., \& Pacholczyk, A. G. 1973, A\&A, 29,
179}

\ref{Rees, M. J., Begelman, M. C., Blandford, R. D., \& Phinney, E. S. 1982,
Nature, 295, 17}

\ref{Shakura, N. I. \& Sunyaev, R. A. 1973, A\&A, 24, 337}

\ref{Shapiro, S. L., Lightman, A. P., \& Eardley, D. M. 1976, ApJ, 204,
187 (SLE)}

\ref{Shaviv, G. \& Wehrse, R. 1986, A\&A, 159, L5}

\ref{Shaviv, G. \& Wehrse, R. 1989, in Theory of Accretion Disks,
eds F. Meyer et al., p419 (Dordrecht: Kluwer)}

\ref{Smak, J. 1984, PASP, 96, 5}

\ref{Spruit, H. C., Matsuda, T., Inoue, M., \& Sawada, K. 1987, MNRAS, 229,
517}

\ref{Stella, L. \& Rosner, R. 1984, ApJ, 277, 312}

\ref{Stepney, S. \& Guilbert, P. W. 1983, MNRAS, 204, 1269}

\ref{Sunyaev, R. A., et al. 1991a, in Proc. 28th Yamada Conf. on Front. X-ray
Astr. (Nagoya, Japan)}

\ref{Sunyaev, R. A., et al. 1991b, A\&A, 247, L29}

\ref{Sunyaev, R. A. \& Titarchuk, L. G. 1980, A\&A, 86, 127}

\ref{Svensson, R. 1982, ApJ, 258, 335}

\ref{Svensson, R. 1984, MNRAS, 209, 175}

\ref{Wallinder, F. H. 1991, A\&A, 249, 107}

\ref{Wandel, A. \& Liang, E. P. 1991, ApJ, 380, 84}

\ref{White, T. R. \& Lightman, A. P. 1989, ApJ, 340, 1024}

\ref{Zdziarski, A. A. 1985, ApJ, 289, 514}

\vfill\eject
\noindent{\bf Figure Captions}
\bigskip\noindent

\bigskip\noindent
Figure 1. The solid line shows at each scaled radius $r$ the scaled
mass accretion rate $\md$ at which $f=0.5$, i.e. the $\dot m$ at which
exactly half the dissipated energy is radiated and half is advected.
The case shown corresponds to a $10M_\odot$ black hole with
$\alpha=0.3$, $\beta=0.5$, and the curve has been obtained by solving
the full equations of \S\S2,3.  The dotted line shows the approximate
relation (4.1) derived under the simplifying assumptions that the ion
and electron temperatures are equal and that bremsstrahlung cooling
dominates.

\bigskip\noindent
Figure 2.  Shows the three solution branches for an accreting
$10M_\odot$ black hole at a radius of $r=10^3$.  The uppermost branch
is the new advection-dominated branch discussed in this paper.  Note
that $f\rightarrow1$ for this branch, corresponding to most of the
dissipated energy being advected with the flow.  The lowermost branch
corresponds to the standard cooling-dominated ($f\rightarrow0$) thin
accretion disk solution.  These two stable branches are connected by
an unstable middle branch, indicated by a dotted line, which
corresponds to the hot solution discovered originally by SLE.  See
Fig. 8 for more details.

\bigskip\noindent
Figure 3. Solid lines in the upper three panels show the critical
accretion rate, $\md_{crit}$, above which the advection-dominated
branch ceases to exist.  Three values of $\alpha$ are considered:
0.03, 0.1, 0.3.  The advection parameter $f$ is $\sim0.3$ on these
lines.  The dotted lines correspond to $f=0.9$ where 90\% of the
dissipated energy is advected.  The bottom three panels show the
variations of the ion temperature $T_i$ and electron temperature $T_e$
along the lines in the upper panels.  Note that the accreting plasma
is single temperature and virial at large radii $r\ \sgreat\ 10^3$.
For $r\ \sles\ 10^3$, $T_i$ remains nearly virial but $T_e$ saturates
because of poor coupling between the ions and the electrons plus the
availability of a variety of efficient cooling mechanisms for the
electrons.

\bigskip\noindent
Figure 4. Comparison of the critical lines $\md_{crit}(r)$ (upper
panel) and the ion and electron temperatures (lower panel) for models
of two accreting black holes of masses $10M_{\sun}$ (solid lines) and
$10^8M_{\sun}$ (dashed lines).  Note that the two models give
virtually identical results, showing that advection-dominated flows
around black holes are essentially mass-independent.

\bigskip\noindent
Figure 5. Similar to Fig. 3 but for an accreting neutron star with
scaled mass $m=1.4$ and radius $r_*=2.5$ (corresponding to a stellar
radius of 10.5 km).  The main new feature in these models is that the
advected energy is assumed to be thermalized and radiated from the
stellar surface.  This radiation provides an important source of
Compton cooling for the accreting gas.  Note that the critical lines
$\md_{crit}$ for these models are significantly lower than in Fig. 3,
and that the electron temperature is also lower.  In panels {\it a}
and {\it b}, the radiation from the neutron star is taken to be a
fully thermalized blackbody, while in panel {\it c} it is assumed to
be at a temperature of $10^9$ K.

\bigskip\noindent
Figure 6. Indicates the relative importance of the various cooling
mechanisms for two black hole models ($10M_\odot$ in panel a and
$10^8M_\odot$ in panel b) and a neutron star model (panel c).
Bremsstrahlung cooling ($q_{br}^-+q_{br,C}^-$) is shown by dotted
lines, synchrotron cooling ($q_{synch}^-+q_{synch,C}^-$) is shown by
short-dashed lines, and Compton cooling by soft stellar photons
($q_{*,C}^-$) is indicated by long-dashed lines.  The various cooling
terms were computed along the critical lines $\md_{crit}(r)$ in Figs.
3--5.  In the black hole models bremsstrahlung dominates at large
radii and synchrotron at small radii.  In the neutron star model,
Comptonization of stellar photons dominates out to $r\sim 10^5$.

\bigskip\noindent
Figure 7. Indicates the zones in the $\md r$ plane of an accreting
$10M_\odot$ black hole where various cooling terms dominate.  Zone 1
corresponds to the region where bremsstrahlung cooling (whose volume
rate is given by $q_{br}^-$) dominates, zone 2 to Comptonized
bremsstrahlung ($q_{br,C}^-$), zone 3 to synchrotron emission
($q_{synch}^-$), and zone 4 to Comptonized synchrotron
($q_{synch,C}^-$).  Note that bremsstrahlung-related emission
dominates at large $r$ and synchrotron-related emission at small $r$,
while Comptonization of these is important at high $\md$.

\bigskip\noindent
Figure 8. (a) Shows the nature of the three solutions, marked A, B, C,
for a typical case consisting of a $10M_\odot$ black hole accreting
with $\md=10^{-3}$.  The situation shown corresponds to a radius
$r=10^3$.  Solution A corresponds to a standard thin accretion disk
solution.  Because $T_i$ increases with increasing $q^-/(1-f)q^+$,
this is a stable solution as discussed in the text.  Solution C is our
new advection-dominated solution and is again stable.  Solution B is,
however, an unstable solution as a result of the wrong sign of the
slope.  Because the two stable solutions A, C bracket the unstable
solution B, the gas can find a thermally stable configuration starting
from any initial configuration.  (b) Similar to panel a, but with
advection ignored.  Note that solution C disappears.  If this gas is
set up with an initial $T_i\ \sgreat\ 10^{7.5}$K, then it will heat up
without limit and will be unable to find a thermally stable
equilibrium.  This paradoxical situation is the result of ignoring
advection.  (c) Similar to Fig. 2, but showing $T_i$ along the
vertical axis instead of $f$.  The branches are labeled A, B, C, as in
panel a.  Note that B and C both correspond to hot solutions.  B is
the usual hot solution discussed in the literature, discovered
originally by SLE, while C is the new advection-dominated hot solution
of this paper.  (d) Another representation, where the vertical axis
shows $\md$ and the horizontal axis shows the surface density
$\Sigma$.  Note the characteristic S-curve.  Note also that, for all
$\md$, there is at least one stable solution available.  Therefore,
this gas will not undergo thermal limit cycle behavior.  The shape of
the S-curve in this diagram is different from that in other situations
where a limit cycle is argued to occur.  If radiation pressure effects
are included (which are ignored in our calculations), then the A
branch will develop another S curve near the top right corner of the
panel (see Abramowicz et al. 1988) because of the Lightman-Eardley
(1974) instability.

\bigskip\noindent
Figure 9.  (a) The dashed line shows the boundary above which the
Lightman-Eardley (1974) instability of a radiation-dominated thin
accretion disk operates.  The case shown corresponds to an accreting
$10M_\odot$ black hole.  The solid line shows $\dot m_{crit}(r)$.  In
the small triangular region between the two lines, the thin disk
solution is unstable but the advection-dominated solution is stable.
Advection-dominated accretion must occur in this region.  (b)
Equivalent results for a $10^8M_\odot$ black hole.

\bigskip\noindent
Figure 10.  (a) The dotted line indicates the track of a $10M_\odot$
black hole accreting at a rate $\md=10^{-2}$.  The solid circle
indicates a critical radius $r_{crit}=10^{5.5}$ such that for
$r<r_{crit}$ the track is located in a region where
advection-dominated accretion is allowed ($\md<\md_{crit}$) whereas
for $r>r_{crit}$ only thin disk accretion is possible
($\md>\md_{crit}$).  If the accretion flow begins with an outer radius
$r_{out}<r_{crit}$ and if it is initially in a nearly virial state,
then it will accrete in an advection-dominated mode all the way down
to the horizon.  (b) Shows a case where $r_{out}>r_{crit}$.  We
propose that the accreting material forms a disk plus a corona with
the indicated mass accretion rates, $\md_{disk}$ and $\md_{cor}$, in
the region $r>r_{crit}$.  For $r<r_{crit}$ we propose that the gas
switches completely to the advection-dominated mode.

\bigskip\noindent
Figure 11.  The solid line in the panel on the left shows the
variation of the radiated luminosity of an accreting $10M_\odot$ black
hole as a function of the mass accretion rate $\dot M$, for
$\alpha=0.3$, $\beta=0.5$.  The dashed line shows the dependence
expected if the radiative efficiency is $\eta=0.1$, i.e. if $L=0.1\dot
M c^2$.  The actual efficiency is lower, especially at low accretion
rates, because the flow is advection-dominated.  The panel on the
right corresponds to an accreting neutron star.  In this case, the net
radiative efficiency is $\eta\sim0.2$ regardless of $\dot M$.
However, most of the energy is advected with the accretion flow and
radiated at the stellar surface.  The luminosity radiated by the
accretion flow itself is shown by the dotted line, which is seen to
have a low efficiency.

\bye